\documentclass[aps, nofootinbib, prd, superscriptaddress, tightenlines,
notitlepage, twocolumn, floatfix]{revtex4-1}
\usepackage{tabularx}
\usepackage{graphicx}
\usepackage{epsfig}
\usepackage{bm}
\usepackage{amssymb,bm,tensor}
\usepackage{textcomp}
\usepackage{xcolor}
\usepackage{amsmath}
\usepackage[varg]{txfonts}
\usepackage{enumerate}
\usepackage{mathtools}
\usepackage[normalem]{ulem}
\usepackage{bm} % bold math
\usepackage[OT1]{fontenc}
\usepackage[colorlinks]{hyperref}
\usepackage{multirow}

\def\al{\alpha}
\def\be{\beta}
\def\ga{\gamma}
\def\de{\delta}
\def\ep{\epsilon}

\def\et{\eta}
\def\th{\theta}

\def\ka{\kappa}
\def\la{\lambda}

\def\rh{\rho}

\def\ph{\phi}
\def\vp{\varphi}
\def\ch{\chi}

\def\om{\omega}
\def\Ga{\Gamma}
\def\De{\Delta}
\def\Th{\Theta}

\def\Ph{\Phi}

\def\Om{\Omega}

\def\mn{{\mu\nu}}

\def\lsim{\mathrel{\rlap{\lower4pt\hbox{\hskip1pt$\sim$}}
    \raise1pt\hbox{$<$}}}
\def\gsim{\mathrel{\rlap{\lower4pt\hbox{\hskip1pt$\sim$}}
    \raise1pt\hbox{$>$}}}
\def\sqr#1#2{{\vcenter{\vbox{\hrule height.#2pt
         \hbox{\vrule width.#2pt height#1pt \kern#1pt
         \vrule width.#2pt}
         \hrule height.#2pt}}}}

\def\prt{\partial}
\def\lrpartial{\raise 1pt\hbox{$\stackrel\leftrightarrow\partial$}}

\def\part2{\partial_\alpha \partial^\alpha}

\def\xx'{|\vec x -\vec x'|}

\def\b2{b^\al b_\al}

\newcommand{\beq}{\begin{equation}}
\newcommand{\eeq}{\end{equation}}
\newcommand{\bea}{\begin{eqnarray}}
\newcommand{\eea}{\end{eqnarray}}
\newcommand{\bit}{\begin{itemize}}
\newcommand{\eit}{\end{itemize}}
\newcommand{\rf}[1]{(\ref{#1})}

\newcommand{\Reply}[1]{#1}  % Reply

\newcommand{\KIAA}{\affiliation{Kavli Institute for Astronomy and
Astrophysics, Peking University, Beijing 100871, China}}
\newcommand{\DoA}{\affiliation{Department of Astronomy, School of Physics,
Peking University, Beijing 100871, China}}

\newcommand{\NAOC}{\affiliation{National Astronomical Observatories,
Chinese Academy of Sciences, Beijing 100012, China}}

\begin{document}
\title{\Reply{Precession of spheroids under Lorentz violation and
observational consequences for neutron stars}}
\author{Rui Xu}\email[Corresponding author: ]{xuru@pku.edu.cn}\KIAA
\author{Yong Gao}\DoA\KIAA
\author{Lijing Shao}\email[Corresponding author: ]{lshao@pku.edu.cn}\KIAA\NAOC
\date{\today}

\begin{abstract}
The Standard-Model Extension (SME) is an effective-field-theoretic
framework that catalogs all Lorentz-violating field operators. The
anisotropic correction from the minimal gravitational SME to Newtonian
gravitational energy for spheroids is studied, and \Reply{the rotation of rigid spheroids} is solved with perturbation method and numerical approach. The
well-known forced precession solution given by Nordtvedt in the
parameterized post-Newtonian formalism is recovered and applied to two observed solitary millisecond pulsars to set bounds on the
coefficients for Lorentz violation in the SME framework. A different
solution, which describes the rotation of an otherwise free-precessing star
in the presence of Lorentz violation, is found, and its consequences on
pulsar signals and continuous gravitational waves (GWs) emitted by neutron
stars (NSs) are investigated. The study provides new possible tests of
Lorentz violation once free-precessing NSs are firmly identified in the
future.
\end{abstract}

\maketitle

%---------------------------------------------------------------------
\section{Introduction}
%---------------------------------------------------------------------

Lorentz invariance, which claims \Reply{the equivalence between any two inertial reference frames in
formulating the laws of physics}, is a fundamental principle in both general
relativity (GR) and the Standard Model of particle physics. However, in
pursuing a unified theory of gravity and quantum particles, violation of
Lorentz invariance was suggested \cite{Kostelecky:1988zi,
Kostelecky:1989jp, Kostelecky:1991ak, Will:1972zz, Will:2018bme}, and has
been treated as a possible suppressed effect emerging from the Planck scale
where the unified theory lives. Therefore, searching for Lorentz-invariance
violation in high-precision terrestrial experiments and astrophysical
observations~\cite{Kostelecky:2008ts, Will:2018bme, Shao:2016ezh} not only
serves as a necessary test of our current best knowledge of the laws of
physics, but also provides us the chance to obtain information on the
underlying theory of quantum gravity that is otherwise unattainable
directly in the experiments and observations with limited energy scales.

A practical framework to study Lorentz violation without digging into the
extensive intricacies of the underlying theory is the Standard-Model
Extension (SME) developed since 1990s \cite{Colladay:1996iz,
Colladay:1998fq, Kostelecky:2003fs}. Aiming at guiding experimental
searches of Lorentz violation, the SME framework is constructed at the
level of effective field theory and treats all possible Lorentz-violating
operators as perturbations on top of GR and the Standard Model. In this
work we are studying effects of Lorentz violation on \Reply{the rotation} of stars.
This belongs to the gravitational sector of the SME framework, and for
simplicity, we only consider the modification to gravity generated by the
so-called minimal gravitational SME \cite{Bailey:2006fd}, given by the
action
\bea
S = \frac{1}{16\pi} \int \sqrt{-g} \, d^4x \, \left( R + k^{\al\be\ga\de} R_{\al\be\ga\de} \right) + S_k + S_m,
\label{action}
\eea
where the Ricci scalar $R$ represents the usual Einstein-Hilbert term for
GR, $R_{\al\be\ga\de}$ is the Riemann tensor, and $k^{\al\be\ga\de}$ is the
tensor field that breaks local Lorentz invariance when it acquires a
nonzero vacuum expectation value in an underlying theory. The action $S_k$
describes the dynamics of the Lorentz-violating field $k^{\al\be\ga\de}$ at
the level of effective field theory. The symmetry-breaking mechanism is
important but generally unspecified~\cite{Kostelecky:2003fs, Bluhm:2004ep}.
One of the central tasks in the gravitational SME is obtaining approximate
but general expressions for the contribution of $S_k$ in the field
equations using geometric properties of the spacetime manifold like the
Bianchi identity and the diffeomorphism invariance \cite{Bailey:2006fd,
Bailey:2014bta}. The last term $S_m$ is the action of conventional matter
and in this work we consider it Lorentz invariant. Note that the
geometrized unit system where $G=c=1$ is used. We will stick to this unit
system except when units are specified explicitly. Another explanation
about the notations in the remaining of the paper is that repeated indices
are summed even when they both are subscripts or superscripts. Also note
that the Greek letters run over spacetime indices while the Latin letters
are restricted to spatial indices only.

The weak field solution of the metric to the field equations obtained from
action \rf{action} is calculated by \citet{Bailey:2006fd}. Especially, as
the starting point of our study, the anisotropic modification to the
Newtonian potential energy between two point particles, $A$ and $B$, is
\bea
\de U = -\frac{1}{2} \bar s^{ij} \frac{ m_A m_B}{|{\boldsymbol{x}}_A - {\boldsymbol{x}}_B|^3} (x_A^i-x_B^i) (x_A^j - x_B^j) ,
\label{pecorrect}
\eea
where $x_A^i$ and $x_B^i$ are the components of the position vectors
${\boldsymbol{x}}_A$ and ${\boldsymbol{x}}_B$. The quantities $\bar s^{ij}$
are defined as the spatial components of
\bea
\bar s^{\al\be} \equiv 2 \left( \bar k^{\al\mu\be\nu} g_{\mn} - \frac{1}{4} g^{\al\be} \bar k^{\la\mu\ka\nu} g_{\mn} g_{\la\ka} \right),
\eea
with $\bar k^{\al\be\ga\de}$ being the vacuum expectation value of the
tensor field $k^{\al\be\ga\de}$~\cite{Kostelecky:2003fs}. In practical
calculations aiming at testing the theoretical predictions against
experimental results, the vacuum expectation value $\bar k^{\al\be\ga\de}$
is taken to be constant around the experimental setting in approximate
inertial frames, and its components, as well as the combinations $\bar
s^{\al\be}$, are called coefficients for Lorentz violation.

Stringent constraints have been set on the SME coefficients for Lorentz
violation using an extensive number of laboratory experiments and
astrophysical observations \cite{Kostelecky:2008ts}. In fact, they include
results from considering effects of Eq.~\rf{pecorrect} on \Reply{the rotation} of the
Sun \cite{Bailey:2006fd} and two isolated millisecond pulsars
\cite{Shao:2014oha}. In this work, \Reply{we are going to consider the two isolated millisecond pulsars in Ref.~\cite{Shao:2014oha} and set new constraints on the coefficients $\bar s^{ij}$.} Our work
complements the results in the literature as we perform a rigorous study on
the rotation of a spheroidal star under the anisotropic gravitational
self-energy caused by Eq.~\rf{pecorrect} to serve as the theoretical basis.
%The immediate achievement is an improvement of the constraints obtained
%from the two isolated pulsars when the maximal-reach approach is
%considered~\cite{Tasson:2019kuw}. 
In addition, our calculation clarifies
the fact that the effect of Lorentz violation adds to stationary spinning
stars as well as free-precessing stars. While the former case is the
Lorentz-violating precession considered in the literature and has been used
to set constraints on the coefficients for Lorentz violation, the latter
case has not been investigated to our knowledge. When Lorentz violation
modifies the rotation of an otherwise free-precessing neutron star (NS),
the pulsar signal and associated continuous gravitational waves (GWs)
emitted by the star~\cite{Gao:2020zcd} will change accordingly. A
significant part of our work is devoted to this new topic.

Before ending the introduction, we are obligated to point out that the
gravitational sector of the SME is not the only framework to study Lorentz
violation in gravity. The celebrated parameterized post-Newtonian (PPN)
formalism also includes coefficients describing possible preferred-frame
effects in metric gravitational theories \cite{Will:2018bme, Will:2014kxa}.
The first study of Lorentz-violating effects on the spins of the Sun and of
millisecond pulsars was done by \citet{Nordtvedt:1987} considering the specific PPN
modification\footnote{Notice that we follow the usage of $\al_2$ by
\citet{Will:2018bme}, related to the one used by \citet{Nordtvedt:1987}
via $\al_2 = 2 \al^{\rm Nordtvedt}_2$. }
\bea
\de U_{\rm PPN} = \frac{\al_2}{2} \frac{ m_A m_B}{|{\boldsymbol{x}}_A - {\boldsymbol{x}}_B|^3} w^{i} w^j (x_A^i-x_B^i) (x_A^j - x_B^j),
\label{ppndeu}
\eea 
to the Newtonian potential. The coefficient $ \al_2$ controls the size of
Lorentz violation while the ``absolute'' velocity $\boldsymbol{w}$ with
repect to the preferred inertial frame picks up a special direction and
breaks Lorentz invariance~\cite{Will:1972zz}. Our study is originally
inspired by Nordtvedt's work~\cite{Nordtvedt:1987}, and our result recovers
his solution when the replacement~\cite{Bailey:2006fd,Shao:2014oha}
\bea
\bar s^{ij} \rightarrow -\al_2 w^i w^j ,
\label{replace}
\eea
is applied and proper approximations are made. The SME framework is more
generic than the PPN formalism in terms of Lorentz
violation~\cite{Bailey:2006fd}. Detailed comparisons will be presented in
relevant paragraphs of the paper.

The organization of the paper is as follows. We start with calculating the
anisotropic gravitational self-energy due to Eq.~\rf{pecorrect} for a
spheroidal star in Sec.~\ref{sec2a}. Then, in Sec.~\ref{sec2b}, we
investigate the solutions to the rotational equations of motion thoroughly,
where both perturbation method and numerical calculation are employed. The
observational consequences are discussed in Sec.~\ref{sec3} with regard to NSs. Section \ref{sec3a} considers stationary spinning stars
affected by Lorentz violation and obtains constraints on the coefficients
for Lorentz violation from observations of two solitary millisecond pulsars. Sections \ref{sec3b} and
\ref{sec3c} consider free-precessing stars affected by Lorentz violation
and provide preliminary signal templates of pulsar pulses and continuous
GWs for fitting observational data in the future. Finally, conclusions are
summarized in Sec.~\ref{sec4}. Appendix~\ref{app1} displays some
expressions for uniform spheroids, which are useful for estimating
numerical values.

%%%%%%%%%%%%%%%%%%%%%%%%%%%%%%%%%%%%%%%%%%%%%%%%%%%%%%%%%%%%%%%%%%%%%%%%%%%%%%
\section{rotation of a spheroid under Lorentz-violating gravity}
\label{sec2}

\subsection{Anisotropic gravitational self-energy}
\label{sec2a}

%----------------------------------------------------------------------
\begin{figure}
    \centering
    \includegraphics[width=6cm]{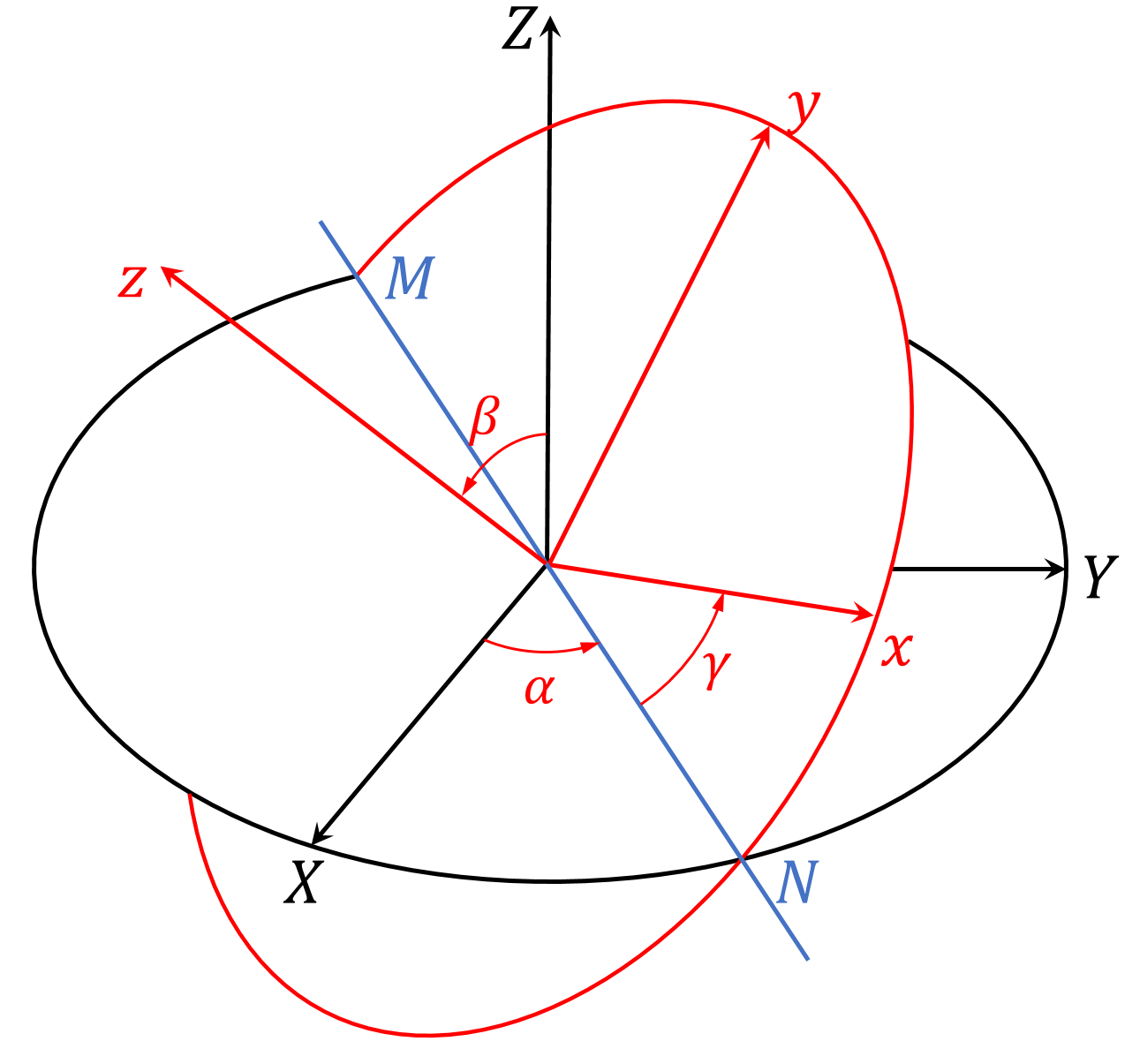}
    \caption{Euler angles $\al, \, \be$ and $\ga$ connecting the
    $X$-$Y$-$Z$ frame and the $x$-$y$-$z$ frame. The line $MN$ is the
    intersection of the $X$-$Y$ plane and the $x$-$y$ plane. }
    \label{euler1}
\end{figure}
%----------------------------------------------------------------------

For a nonspherical star, the integral of the potential energy correction in
Eq.~\rf{pecorrect} depends on the orientation of the star, causing a torque
during its rotation. Specifically speaking, we calculate $\de U$ in the
body frame of the star at any instant,
\bea
\de U = -\frac{1}{4} \bar s^{ij} \int d^3x \, d^3x' \rh(\boldsymbol{x}) \rh(\boldsymbol{x'}) \frac{ \left( x^i - x^{\prime\,i} \right) \left( x^j - x^{\prime\,j} \right) }{|\boldsymbol{x} - \boldsymbol{x'}|^3} ,
\label{pecorrect1}
\eea 
where $\rh$ is the density of the star, and is assumed to be independent of
time in the body frame. The orientation dependence goes into $\de U$ as the
star rotates. If we set up an inertial frame $X$-$Y$-$Z$, then $\bar
s^{ij}$ $(i,j=x,y,z)$ in the body frame is related to $\bar s^{IJ}$
$(I,J=X,Y,Z)$ in the inertial frame by a rotation transformation
\bea
\bar s^{ij} = R^{iI} R^{jJ} \bar s^{IJ} .
\eea 
Noticing that $\bar s^{IJ}$ are the constant coefficients for Lorentz
violation, $\de U$ therefore depends on the orientation of the body through
the rotation matrix $R^{iI}$. The orientation of the body is conveniently
described by the Euler angles (see Fig. \ref{euler1}), which are the
kinematic quantities to be solved from the equations of motion. For later
use, we write the elements of the rotation matrix as
\bea
R^{iI} = \hat {\boldsymbol{e}}_i \cdot \hat {\boldsymbol{e}}_I ,
\label{rmat}
\eea 
where $\{ \hat {\boldsymbol{e}}_i \}$ $(i=x,y,z)$ is the basis of the body
frame $x$-$y$-$z$, and $ \{ \hat {\boldsymbol{e}}_I \}$ $(I=X,Y,Z)$ is the
basis of the inertial frame $X$-$Y$-$Z$. The inner products can be computed
using the relations~\cite{landau1960course}
\bea
\hat {\boldsymbol{e}}_x &=& \left(\cos\al\cos\ga-\sin\al\cos\be\sin\ga \right) \hat {\boldsymbol{e}}_X 
\nonumber \\
 && + \left(\sin\al\cos\ga+\cos\al\cos\be\sin\ga \right) \hat {\boldsymbol{e}}_Y + \sin\be\sin\ga \,\hat {\boldsymbol{e}}_Z ,
\nonumber \\
\hat {\boldsymbol{e}}_y &=& \left(-\cos\al\sin\ga-\sin\al\cos\be\cos\ga \right) \hat {\boldsymbol{e}}_X 
\nonumber \\
 && + \left(-\sin\al\sin\ga+\cos\al\cos\be\cos\ga \right) \hat {\boldsymbol{e}}_Y + \sin\be\cos\ga \,\hat {\boldsymbol{e}}_Z ,
\nonumber \\
\hat {\boldsymbol{e}}_z &=& \sin\al\sin\be \,\hat {\boldsymbol{e}}_X - \cos\al\sin\be \,\hat {\boldsymbol{e}}_Y + \cos\be \,\hat {\boldsymbol{e}}_Z .
\label{trans}
\eea 

Back to the potential energy correction in Eq.~\rf{pecorrect1}, to proceed
the calculation analytically as much as possible, we focus on the simple
case where the star is a spheroid described by
\bea
\frac{x^{2} + y^{2}}{a_1^2} + \frac{z^{2}}{a_3^2}  = 1 ,
\label{spheroid}
\eea
in the body frame and its density $\rh$ is axisymmetric about the $z$-axis (for an ellipsoid, see e.g., Ref.~\cite{Xu:2021dcw}). In such case, one can show that $\bar
s^{xy}, \, \bar s^{xz}$ and $\bar s^{yz}$ do not appear in $\de U$, and
that $\bar s^{xx}$ and $\bar s^{yy}$ appear as the combination $\bar s^{xx}
+ \bar s^{yy}$. Because the trace $\bar s^{xx} + \bar s^{yy} + \bar s^{zz}$
is rotationally invariant and hence does not contribute to the anisotropic
correction, the contribution from $\bar s^{xx} + \bar s^{yy}$ can be
represented by $\bar s^{zz}$ so that the true anisotropic correction of the
potential energy is simply
\bea
\de U = C \, \bar s^{zz}, 
\label{pecorrect2}
\eea
with the constant $C$ being
\bea
C = \frac{1}{4} \int d^3x \, d^3x' \rh(\boldsymbol{x}) \rh(\boldsymbol{x'}) \frac{ \left( x - x' \right)^2 - \left( z - z' \right)^2 }{|\boldsymbol{x} - \boldsymbol{x'}|^3} .
\eea

The result \rf{pecorrect2} deserves a comparison with that of
\citet{Nordtvedt:1987}. Treating the potential correction \rf{ppndeu} as a
perturbation on top of the Newtonian potential, \citet{Nordtvedt:1987} used
the tensor virial relation to obtain the anisotropic gravitational
self-energy for a spheroid star as
\bea
\de U_{\rm PPN} = -\frac{1}{2} \al_2 T_{\rm rot} ( \boldsymbol{w} \cdot \hat{\boldsymbol{\Om}} )^2 , 
\label{ppndeu1}
\eea
where $T_{\rm rot}$ is the rotational kinetic energy and
$\hat{\boldsymbol{\Om}}$ is the unit vector along the angular velocity of
the star. 
%Note that the original expression of $\de U_{\rm PPN}$ in
%Ref.~\cite{Nordtvedt:1987} has a plus sign which is incorrect. 

First of all, we point out that with the replacement \rf{replace},
Eq.~\rf{pecorrect2} recovers the form of Eq.~\rf{ppndeu1} when the star
spins stationarily along the $z$-axis. As Nordtvedt obtained, for
nonrelativistic stars, the constant $C$ is equal to $T_{\rm rot}/2$ by
virtue of the tensor virial relation
\bea
K^{ij} + \de^{ij} P - U^{ij} = 0,
\eea
where $\de^{ij}$ is the Kronecker delta, and
\bea
K^{ij} &\equiv& \int \rh v^i v^j d^3x,
\nonumber \\
P &\equiv& \int p d^3x ,
\nonumber \\
U^{ij} &\equiv& \int \rh x^i \prt_j \Phi d^3x,  
\eea
with ${\boldsymbol{v}}$ being the velocity field inside the star, $p$ being
the pressure inside the star, and $\Phi$ being the usual Newtonian
potential
\bea
\Phi ({\boldsymbol{x}}) = -\int d^3x' \frac{ \rh(\boldsymbol{x'}) }{|\boldsymbol{x} - \boldsymbol{x'}|} .
\label{newp}
\eea
The result $C = T_{\rm rot}/2$ is straightforward to prove once we notice
the integral identity
\bea
U^{ij} = \frac{1}{2} \int d^3x \, d^3x' \rh(\boldsymbol{x}) \rh(\boldsymbol{x'}) \frac{ \left( x^i - x^{\prime\,i} \right) \left( x^j - x^{\prime\,j} \right) }{|\boldsymbol{x} - \boldsymbol{x'}|^3},
\eea
and for a star spinning along the $z$-direction,
\bea
K^{zz} = 0, \quad T_{\rm rot} = \frac{1}{2} \left( K^{xx} + K^{yy} \right) .
\eea  
In Appendix~\ref{app1}, the constant $C$ is exhibited for uniform spheroids
by utilizing the Newtonian potential inside an ellipsoid of uniform density
(see e.g., Refs.~\cite{Poisson:2014, 1962ApJ...136.1037C}).

Secondly, we notice that Eq.~\rf{pecorrect2} and Eq.~\rf{ppndeu1} are
different for a generally rotating spheroidal star whose angular velocity
is not aligned with its symmetric axis. The origin of the difference is
that Eq.~\rf{ppndeu1} requires the star to be a stationary fluid in
equilibrium while Eq.~\rf{pecorrect2} applies as long as the star has an
axisymmetric mass distribution. Whether the star is a fluid or a rigid body
does not affect the validity of Eq.~\rf{pecorrect2}. Now comes the
question: How could the state of the star influence the alignment between
its angular velocity and symmetric axis? In fact, if the star is a rigid
body, then the direction of its angular velocity can be different from
its symmetric axis. But when it is a fluid, the deformation is caused by
rotation, and therefore its symmetric axis must be aligned with its angular
velocity as long as the system is in equilibrium.

To better understand the context and the limitation for Eq.~\rf{ppndeu1},
let us consider a freely rotating star, namely when there is no torque. If
modeled as a rigid body, depending on whether the angular velocity is
aligned with the symmetric axis, two kinds of solution exist: the
stationary spinning solution where the star spins around the fixed
symmetric axis, and the free-precessing solution where the star spins
around its symmetric axis while that axis rotates around the fixed
direction of the conserved angular momentum~\cite{landau1960course}. If
modeled as a fluid, the free-precessing solution cannot exist because even
if the angular velocity is not aligned with the symmetric axis initially,
the deformation caused by rotation will eventually change the symmetric
axis to the direction of angular velocity, leaving the star in the state of
stationary spin.

When the torque caused by Lorentz violation is taken into consideration,
the two types of solution for a freely rotating star, namely the stationary
spinning solution and the free-precessing solution, might be treated as the
zeroth-order solution for applying perturbation method. The details are
discussed in Sec.~\ref{sec2b}. Here we just point out that the effect of
the Lorentz-violating torque is forcing the star to precess around one of
the principal axes of the $\bar s^{ij}$ tensor. When this effect is added
to a stationary spinning star, we call the motion {\it{single
precession}}. When the forced precession due to Lorentz violation is added
to a free-precessing star, we call the motion {\it{twofold precession}}.

In the single-precession solution, the precession is caused by Lorentz
violation, which is assumed to be small. Therefore, the total angular
velocity can be approximated as aligned with the symmetric axis of the
star. This enables us to put Eq.~\rf{pecorrect2} in the form of
Eq.~\rf{ppndeu1} at the leading order of Lorentz violation via the
replacement \rf{replace}. So the single-precession solution is the one
found by \citet{Nordtvedt:1987}. However, for our result,
there is no need to restrict it to fluid stars as Eq.~\rf{pecorrect2}
applies to fluids as well as rigid bodies. The other solution that we
defined, the twofold-precession solution, as its name suggests, is a
superposition of the free precession where the symmetric axis of the star
precesses around the angular momentum, and the forced precession where the
angular momentum precesses around one of the principal axes of the $\bar
s^{ij}$ tensor. Though only applicable to stars if they are modeled as rigid
bodies, the twofold-precession solution describes a new Lorentz-violating
effect in the rotation of stars that has not been studied in the literature
to our knowledge.

Before moving to study the solutions in detail, \Reply{a brief clarification on
the model of NS deformation might be useful.} We will adopt the
model described by \citet{2001MNRAS.324..811J} where the deformation of the
star has two contributions: the centrifugal deformation and the Coulomb
deformation. The centrifugal deformation is the fluid deformation that
scales with the square of the angular velocity, while the Coulomb
deformation describes the rigid deviation from the spherical shape
sustained by the electrostatic force. The fact that the electrostatic
interaction is much weaker than the gravitational interaction in NSs
implies that the rigid Coulomb deformation is much smaller than the fluid
centrifugal deformation. Using the oblateness defined as
\bea
\ep = \frac{ I^{zz} - I^{xx} }{ I^{xx} } ,
\label{obl}
\eea
where $I^{xx}$ and $I^{zz}$ are the eigenvalues of the moment of inertia tensor,
to characterize the deformations of spheroidal stars, an estimation for the centrifugal
deformation is
\bea
\hspace{-0.5cm}
\ep_{f} \approx \frac{|{\boldsymbol{\Om}}|^2 R^3}{M} \approx 2.1 \times 10^{-3} \left( \frac{{|{\boldsymbol{\Om}}|}/{2\pi} }{100\,{\rm Hz} } \right)^2 \left( \frac{R}{10 \, {\rm km}} \right)^3 \left( \frac{1.4M_{\odot}}{M} \right) ,
\label{epf}
\eea 
where ${\boldsymbol{\Om}}$ is the angular velocity, $R$ is the radius, and
$M$ is the mass of the NS, while the oblateness $\ep_r$ caused by the
Coulomb deformation is about five orders of magnitude smaller
\cite{2001MNRAS.324..811J}. Therefore, when applying the single-precession
solution to NSs, the rigid deformation $\ep_r$ can be neglected and the
deformation is described by the fluid oblateness $\ep_f$. However, the fact
that NSs can have rigid deformation described by $\ep_r$ is vital when the
twofold-precession solution is to apply. Similar to the free-precessing NSs
considered in Refs.~\cite{Zimmermann:1979ip, 2001MNRAS.324..811J,
Gao:2020zcd, Gao:2020dwy}, NSs with Lorentz-violating gravity under the
twofold precession produce modulated pulsar signals and in the meantime are
sources of continuous GWs, providing potential new tests of Lorentz
violation. This will be the central topic in Sec.~\ref{sec3}.

Finally, we point out that Lorentz violation itself also deforms
stars~\cite{Altschul:2006uu, Xu:2019gua}. But then the deformation due to Lorentz violation
contributes at the second order in terms of the coefficients for Lorentz
violation to the anisotropic gravitational self-energy. To keep the
analysis clear and tractable, we neglect any Lorentz-violating correction
to the structure of stars in this work.

\subsection{Rotation of a spheroidal star}
\label{sec2b}

We are ready to write down the equations of motion and solve the rotation
of a spheroidal star. First, the coefficients for Lorentz violation
naturally fix a convenient inertial frame, namely the one that diagonalizes
the $\bar s^{IJ}$ matrix. We will use it as the inertial frame $X$-$Y$-$Z$. Please note that the inertial frame widely used in the literature is the canonical Sun-centered
celestial-equatorial frame~\cite{Kostelecky:2002hh}. It is generally different from the frame used here as the off-diagonal coefficients for Lorentz violation unlikely happen to vanish in the canonical Sun-centered celestial-equatorial frame.

With the rotation matrix~\rf{rmat}, the anisotropic
gravitational self-energy \rf{pecorrect2} is
\bea
\de U = C \left( (\bar s^{XX} \sin^2\al + \bar s^{YY} \cos^2\al) \sin^2\be + \bar s^{ZZ} \cos^2\be \right) ,
\label{anisopot}
\eea
where $\bar s^{XX}, \, \bar
s^{YY}$, and $\bar s^{ZZ}$ are the eigenvalues of the $\bar s^{IJ}$ matrix.
Then, to express the rotational kinetic energy in terms of the Euler angles
and their derivatives, we employ the kinematic relation between the
velocity components in the body frame and the Euler
angles~\cite{landau1960course},
\bea
\Om^x &=& \dot \al \sin\be \sin\ga + \dot \be \cos\ga ,
\nonumber \\
\Om^y &=& \dot \al \sin\be \cos\ga - \dot \be \sin\ga ,
\nonumber \\
\Om^z &=& \dot \al \cos\be + \dot \ga,
\label{angvel1}
\eea 
where dot denotes time derivative. With $I^{xx} = I^{yy}$, the rotational
kinetic energy simplifies to
\bea
T_{\rm rot} = \frac{1}{2} I^{xx} \left( \dot \al^2 \sin^2\be + \dot \be^2 \right) + \frac{1}{2} I^{zz} \left( \dot \al\cos\be + \dot \ga \right)^2 .
\eea
The Euler-Lagrange equations from $L = T_{\rm rot} - \de U$ can be derived,
and there are two first integrals,
\bea
E &=& T_{\rm rot} + \de U  ,
\nonumber \\
\Om^z &=& \dot \al \cos\be + \dot \ga .
\label{firstint}
\eea 
The total energy $E$ is not much useful in helping simplify the other
equations. The $z$-component of the angular velocity, $\Om^z$, can be used
to eliminate $\dot \ga$ from the other two Euler-Lagrange equations so that
they become
\bea
\sin^2\be \, \ddot \al + \sin{2\be} \, \dot \al \dot \be - \frac{I^{zz} \Om^z }{I^{xx}} \sin\be \, \dot \be &=& - \frac{1}{I^{xx}} \frac{\prt \de U}{\prt \al}, 
\nonumber \\
\ddot \be - \frac{1}{2} \sin{2\be} \, \dot \al^2 + \frac{I^{zz} \Om^z }{I^{xx}} \sin\be \, \dot \al &=& - \frac{1}{I^{xx}} \frac{\prt\de U}{\prt \be} .
\label{eleq}
\eea

First, we consider an illustrative case where only $\bar s^{ZZ}$ is
nonzero. This is the case where the tensor $\bar s^{ij}$ degenerates
to a vector as in the correspondence \rf{replace}. With $\bar s^{XX} = \bar
s^{YY} = 0$, $\de U$ is independent of $\al$ so that solutions with $\ddot
\al = 0, \, \dot \be = 0$ are consistent with Eqs.~\rf{eleq} given
\bea
\dot \al = \frac{I^{zz} \Om^z \pm \sqrt{\left( I^{zz} \Om^z \right)^2 - 8 \bar s^{ZZ} C I^{xx} \cos^2\be } }{2 I^{xx} \cos\be } .
\label{sol1}
\eea 
When $\bar s^{ZZ} = 0$, the ``$+$'' sign recovers the
free-precessing solution and the ``$-$'' sign gives the stationary spinning
solution. Correspondingly, for $\bar s^{ZZ} \ne 0$, the ``$+$'' sign gives
what we call {\it{twofold precession}}, and the ``$-$'' sign gives what we
call {\it{single precession}}. The single-precession solution, once
approximated at the leading order of $\bar s^{ZZ} \leftrightarrow - \al_2
|\boldsymbol{w}|^2$, is the one obtained by \citet{Nordtvedt:1987}.

The solutions \rf{sol1} hold when the matrix $\bar s^{IJ}$ has only one
nonvanishing eigenvalue whose eigenvector is chosen to be the $Z$-axis.
Similar solutions exist when the matrix $\bar s^{IJ}$ has two same
eigenvalues whose eigenvectors are chosen to be the $X$-axis and the
$Y$-axis, because in such case $\de U$ is also independent of $\al$. In
fact, only one nonvanishing eigenvalue is a special case of two same
eigenvalues where the same eigenvalues are zero. In conclusion, when the
matrix $\bar s^{IJ}$ has two same eigenvalues, we can set up the
$X$-$Y$-$Z$ frame properly so that solutions similar to Eq.~\rf{sol1} with
a constant $\be$ exist.

Now we can discuss the solutions in general when the matrix $\bar s^{IJ}$
has three different eigenvalues. In such case, $\de U$ depends on both
$\al$ and $\be$. Solutions with a constant $\be$ do not exist, but as
Lorentz violation is supposed to be small, we use the ansatz
\bea
\al &=& \al_0 + a t + \al^{(1)} ,
\nonumber \\
\be &=& \be_0 + \be^{(1)},
\label{ansatz}
\eea
to find perturbative solutions at the leading order of Lorentz violation.
In the ansatz \rf{ansatz}, $\al_0$ and $\be_0$ are constants, while
$\al^{(1)}$ and $\be^{(1)}$ are functions of time assumed to be much
smaller than 1. The constant $a$, representing the precessing angular
velocity for \Reply{free rotation}, takes the values
\bea
a = \begin{cases}
\frac{\om}{\cos\be_0}, & {\rm free \ precession} , \\
0 ,  & {\rm stationary \ spin} ,
\end{cases} 
\eea
with $\om \equiv I^{zz} \Om^z /I^{xx}$ according to Eq.~\rf{sol1} when
$\bar s^{ZZ} = 0$. Substituting derivatives of the ansatz \rf{ansatz} into
Eqs.~\rf{eleq}, and approximating $\al$ and $\be$ in the equations with
$\al_0+at$ and $\be_0$, we find
\bea
\ddot \al^{(1)} \pm \frac{\om}{\sin\be_0} \, \dot \be^{(1)} &=& - \frac{1}{I^{xx} \sin^2\be_0} \frac{\prt \de U}{\prt \al} \Big|_{\al=\al_0+at,\, \be=\be_0} ,
\nonumber \\
\ddot \be^{(1)} \mp \om\sin\be_0 \, \dot \al^{(1)} &=& - \frac{1}{I^{xx} } \frac{\prt \de U}{\prt \be} \Big|_{\al=\al_0+at,\, \be=\be_0} .
\label{eleq1}
\eea    
Note that the upper signs are for the perturbation to a free-precessing
star, generating twofold-precession solutions, while the lower signs are
for the perturbation to a stationary spinning star, generating
single-precession solutions. We will stick to this sign convention when we
write expressions containing upper and lower signs. Equations \rf{eleq1}
are two coupled oscillation equations for $\dot \al^{(1)}$ and $\dot
\be^{(1)}$ with driven forces. Using Eq. \rf{anisopot} to write out the
right-hand sides of Eqs.~\rf{eleq1}, the solutions can be found as
\bea
\dot \al^{(1)} &=& \mp a^{(1)} \mp \frac{A}{\sin\be_0} \sin{(\om t + \vp)} + \et_\al \, \tilde C \cos{2(at+\al_0)} ,
\nonumber \\
\dot \be^{(1)} &=&  A \cos{(\om t + \vp)} + \et_\be \, \tilde C \sin{2(at + \al_0)} ,
\label{sol2}
\eea
with $A$ and $\vp$ being two integral constants of the homogeneous
solutions. The constants $a^{(1)}, \, \tilde C, \, \et_\al$, and $\et_\be$
are
\bea
a^{(1)} &=& \frac{C \left( 2 \bar s^{ZZ} - \bar s^{XX} - \bar s^{YY} \right) \cos\be_0}{I^{zz} \Om^z} ,
\nonumber \\
\tilde C &=& \frac{C \left( \bar s^{XX} - \bar s^{YY} \right) }{I^{zz} \Om^z} ,
\nonumber \\
\et_\al &=& \begin{cases}
\frac{\cos\be_0 \left(2+\cos^2\be_0 \right)}{4-\cos^2\be_0} , & {\rm twofold \ precession}, \\
\cos\be_0 , & {\rm single \ precession},
\end{cases}
\nonumber \\
\et_\be &=& \begin{cases}
\frac{3\sin\be_0 \cos^2\be_0}{4-\cos^2\be_0} , & ~~~~{\rm twofold \ precession}, \\
\sin\be_0 , & ~~~~{\rm single \ precession} .
\end{cases}
\label{consdef}
\eea
We point out that when $\bar s^{XX} = \bar s^{YY} = 0$, the inhomogeneous
solutions in Eqs.~\rf{sol2} recover the solutions \rf{sol1} at the leading
order of Lorentz violation.

The constant $a^{(1)}$, generated by the constant term on the right-hand
side of the second equation in Eqs.~\rf{eleq1}, represents the forced precession
due to Lorentz violation. It can be absorbed into $a$ by redefining $a$ to
be
\bea
a = \begin{cases}
\frac{\om}{\cos\be_0} - a^{(1)}, & {\rm twofold \ precession} , \\
a^{(1)}, & {\rm single \ precession}  .
\end{cases}
\label{precangvel} 
\eea
Note that the forced precession acts oppositely on a free-precessing star
and a stationary spinning star.

Now we can use this new $a$ in solutions \rf{sol2}, and the benefit is that
the second term in $\dot \be^{(1)}$ is no longer constant for the
single-precession solution. As $\be$ at most changes from $0$ to $\pi$ by
definition, its rate of change really should not contain any constant term.
Keeping in mind that the definition \rf{precangvel} is used, then the
ansatz \rf{ansatz}, together with the solutions for $\al^{(1)}$ and
$\be^{(1)}$,
\bea
\al^{(1)} &=& \pm \frac{A}{\om \sin\be_0} \cos{(\om t + \vp)} + \et_\al \frac{\tilde C}{2a} \sin{2(at+\al_0)} ,
\nonumber \\
\be^{(1)} &=& \frac{A}{\om}\sin{(\om t + \vp)} - \et_\be \frac{\tilde C}{2a} \cos{2(at + \al_0)} ,
\label{sol3}
\eea
gives the perturbation solutions with integral constants $A , \, \vp, \,
\al_0$, and $\be_0$.

%----------------------------------------------------------------------
\begin{figure*}
    \centering
    \includegraphics[width=17cm]{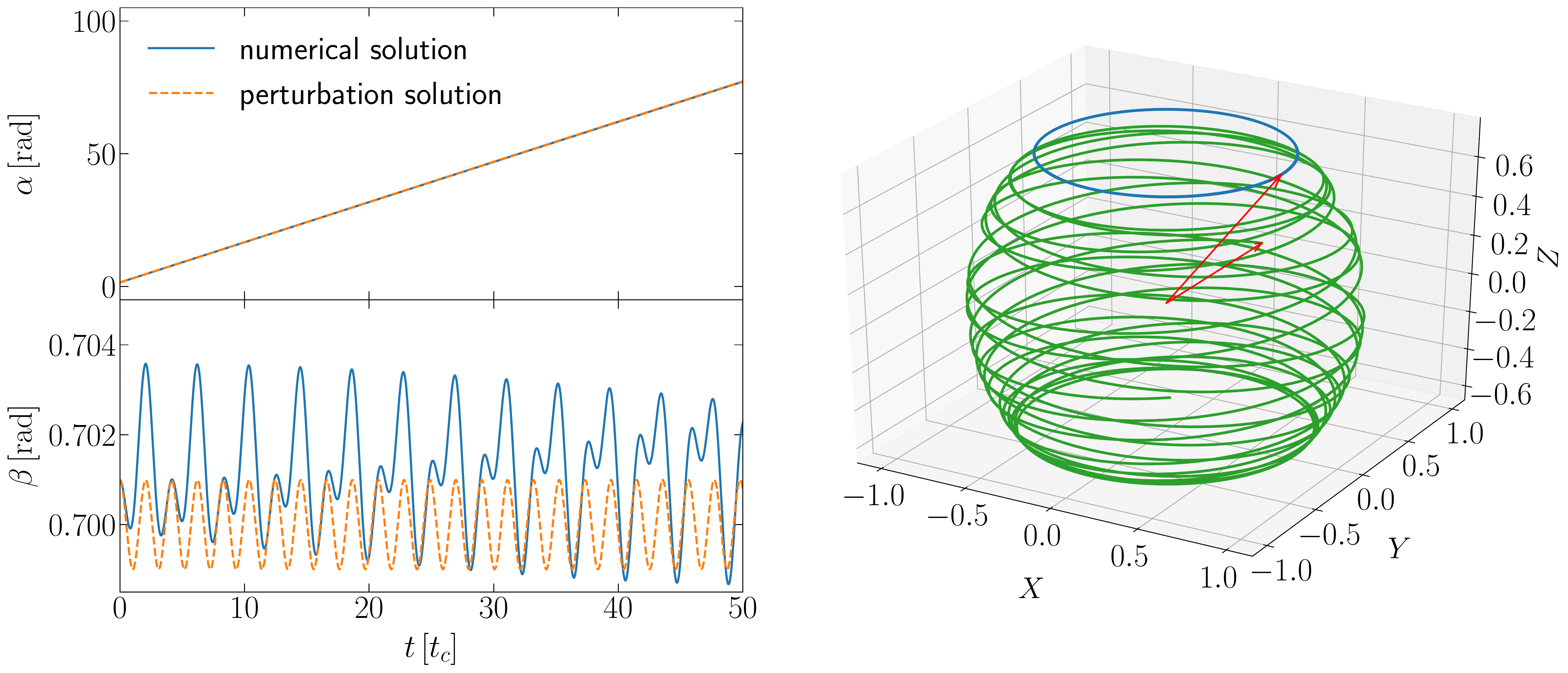}
    \caption{Example I of twofold precession (denoted as tp-I hereafter).
    The left plots are the solutions of the Euler angles $\al$ and $\be$ as
    functions of time. The right plot shows the trajectories of the heads
    of $\hat {\boldsymbol{e}}_x$ (green) and $\hat {\boldsymbol{e}}_z$
    (blue) in the $X$-$Y$-$Z$ frame while their tails are fixed at the
    origin. The arrows mark the two vectors at $t=0$. The initial values,
    $\al|_{t=0}=\frac{\pi}{2}$, $\be|_{t=0} \approx 0.701$, $\dot
    \al|_{t=0} \approx 1.509$, and $\dot \be|_{t=0} = 0$ are adopted by
    setting $\al_0=\frac{\pi}{2}$, $\be_0 = 0.7$, $A=0$, and
    $\vp=\frac{\pi}{2}$ in Eqs.~\rf{integraltoinitial}. Other parameters
    used are $\ga|_{t=0} = 0$, $\Om^z = 1$, $I^{zz}/I^{xx} = 1.1$, and
    $\left\{ \bar s^{XX}, \bar s^{YY}, \bar s^{ZZ} \right\} = \{ 0.02,
    0.01, -0.04\}$. Time and time derivatives are dimensionless under the
    time unit $t_c$ defined in Eq.~\rf{unitt}. }
    \label{fig2}
\end{figure*}
%----------------------------------------------------------------------
\begin{figure*}
    \centering
    \includegraphics[width=17cm]{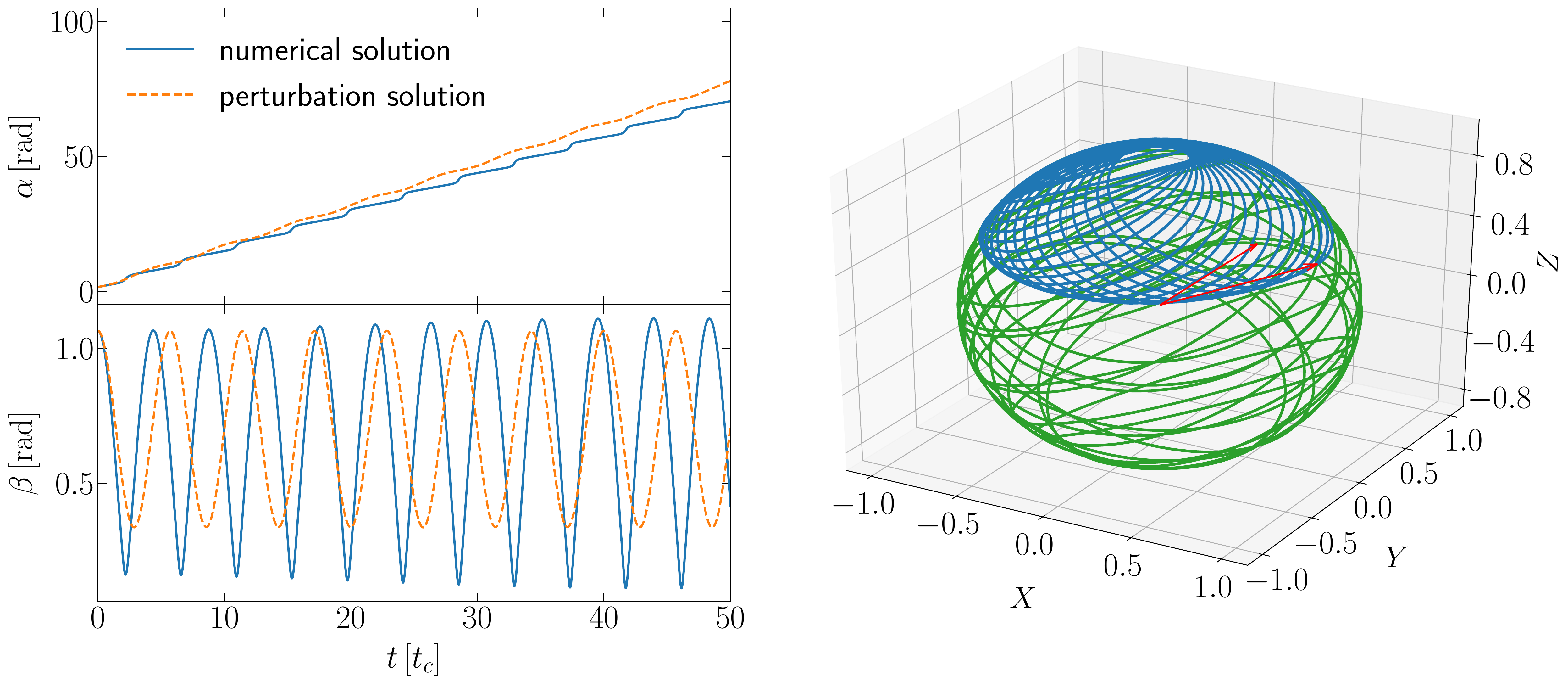}
    \caption{Example II of twofold precession (denoted as tp-II
    hereafter). See the caption of Fig.~\ref{fig2} for the meaning of
    illustration. The initial values, $\al|_{t=0}=\frac{\pi}{2}$,
    $\be|_{t=0} \approx 1.065$, $\dot \al|_{t=0} \approx 0.889$, and $\dot
    \be|_{t=0} = 0$ are adopted by setting $\al_0=\frac{\pi}{2}$, $\be_0 =
    0.7$, $A=0.4$, and $\vp=\frac{\pi}{2}$ in Eqs.~\rf{integraltoinitial}.
    Other parameters are the same as those in Fig.~\ref{fig2}.}
    \label{fig3}
\end{figure*}
%----------------------------------------------------------------------
\begin{figure*}
    \centering
    \includegraphics[width=17cm]{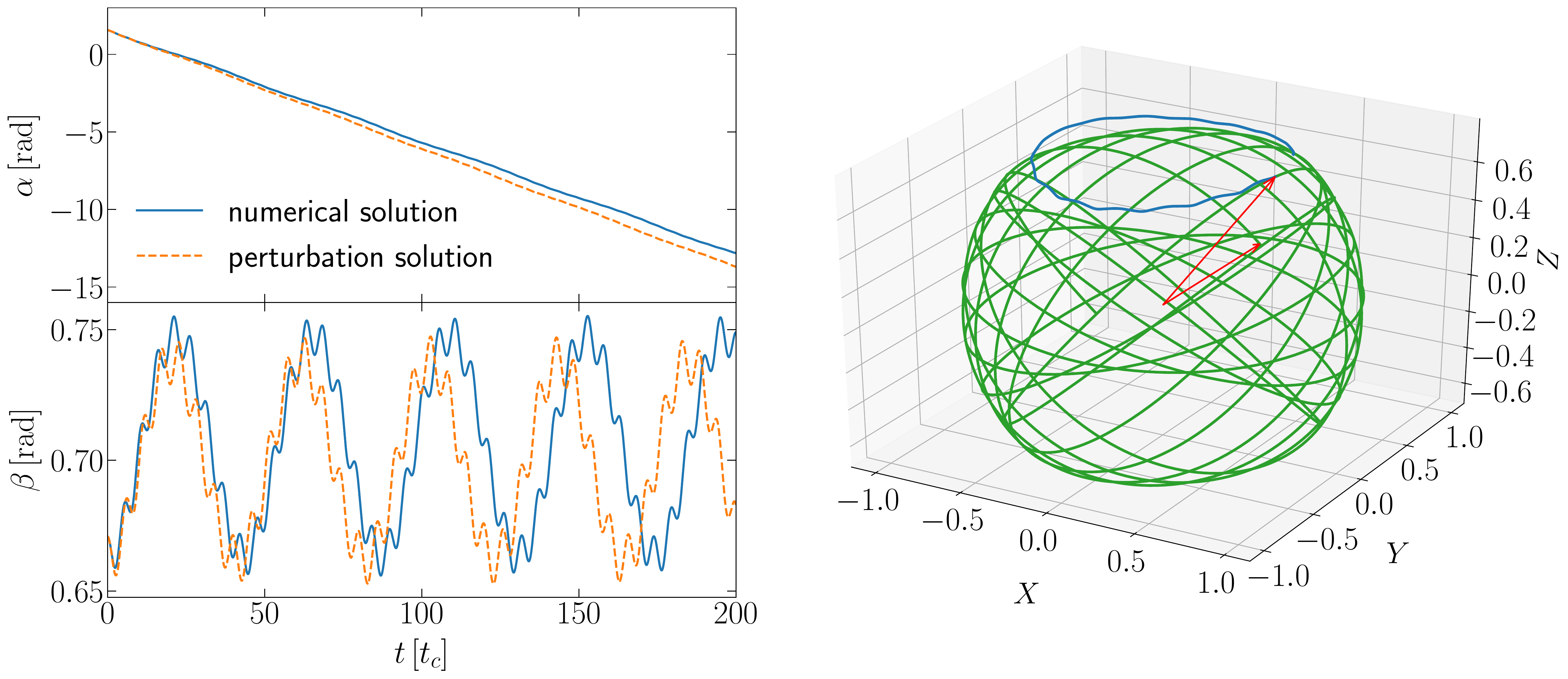}
    \caption{Example I of single precession. See the caption of
    Fig.~\ref{fig2} for the meaning of illustration. The initial values,
    $\al|_{t=0}=\frac{\pi}{2}$, $\be|_{t=0} \approx 0.671$, $\dot
    \al|_{t=0} \approx -0.068$, and $\dot \be|_{t=0} = 0$ are adopted by
    setting $\al_0=\frac{\pi}{2}$, $\be_0 = 0.7$, $A=0.01$, and
    $\vp=\frac{\pi}{2}$ in Eqs.~\rf{integraltoinitial}. Other parameters
    are the same as those in Fig.~\ref{fig2}. }
    \label{fig4}
\end{figure*}
%----------------------------------------------------------------------
\begin{figure*}
    \centering
    \includegraphics[width=17cm]{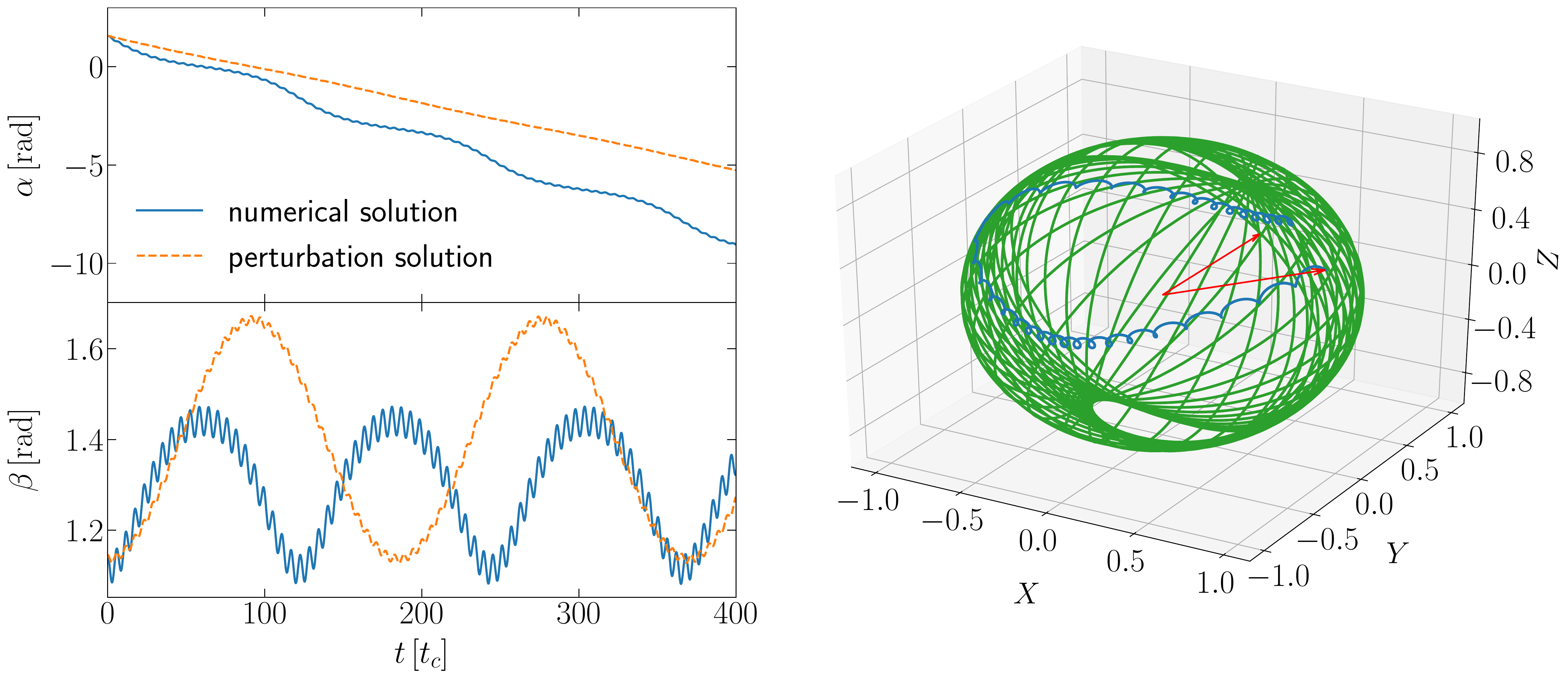}
    \caption{Example II of single precession. See the caption of
    Fig.~\ref{fig2} for the meaning of illustration. The initial values,
    $\al|_{t=0}=\frac{\pi}{2}$, $\be|_{t=0} \approx 1.146$, $\dot
    \al|_{t=0} \approx -0.008$, and $\dot \be|_{t=0} = 0$ are adopted by
    setting $\al_0=\frac{\pi}{2}$, $\be_0 = 1.4$, $A=0.01$, and
    $\vp=\frac{\pi}{2}$ in Eqs.~\rf{integraltoinitial}. Other parameters
    are the same as those in Fig.~\ref{fig2}. }
    \label{fig5}
\end{figure*}
%----------------------------------------------------------------------

The perturbation solutions are useful in analytically illustrating how
Lorentz violation described by the matrix $\bar s^{IJ}$ with three
different eigenvalues affects a freely rotating spheroidal star. But a
complete discussion of the rotational motion has to involve numerical
solutions, because there are solutions to Eqs.~\rf{eleq} unable to be
described by the perturbation solutions even at the leading order of
Lorentz violation. To see this, let us consider solving Eqs.~\rf{eleq} with
initial values given by
\bea
\al|_{t=0} &=& \al_0 \pm \frac{A}{\om \sin\be_0} \cos{ \vp} + \et_\al \frac{\tilde C}{2a} \sin{2\al_0} ,
\nonumber \\
\be|_{t=0} &=& \be_0 + \frac{A}{\om}\sin{\vp} - \et_\be \frac{\tilde C}{2a} \cos{2\al_0} ,
\nonumber \\
\dot \al|_{t=0} &=& a \mp \frac{A}{\sin\be_0} \sin{\vp} + \et_\al \tilde C \cos{2\al_0} , 
\nonumber \\
\dot \be|_{t=0} &=& A \cos\vp + \et_\be \tilde C \sin{2\al_0} .
\label{integraltoinitial}
\eea 
The fact that the initial values $\al|_{t=0}$, $\be|_{t=0}$, $\dot
\al|_{t=0}$, and $\dot \be|_{t=0}$ might be arbitrarily assigned indicates
that the corresponding integral constants $A, \, \vp, \, \al_0$, and $\be_0$
should also be able to take any values. However, the perturbative approach
restricts $A$ to be small, losing solutions where $\dot \al|_{t=0}$ is at
the order of $\om/\cos\be_0$ but significantly different from it. These
solutions belong to the case of twofold precession, but because the
directions of angular momentum around which \Reply{free precession} happens
deviate from the $Z$-axis too much, \Reply{free precession} around the
$Z$-axis is no longer able to serve as the zeroth-order solution for the
perturbation method. More seriously, the perturbative approach assumes
$\al^{(1)}$ and $\be^{(1)}$ to be much smaller than 1. But for the
single-precession solution in Eqs.~\rf{sol3}, this assumption is very much
questionable as a small quantity $a = a^{(1)}$ appears in the denominators
of the second terms of $\al^{(1)}$ and $\be^{(1)}$. Somehow the
perturbative \Reply{single-precession solution} does mimic numerical solutions in
one period of the forced precession when $\be_0$ is small ($\be_0 \lsim
1$). But when $\be_0$ gets close to $\pi/2$, the approximation fails
completely.

Figures \ref{fig2}--\ref{fig5} display examples to compare the perturbation
solutions with numerical solutions, and also illustrate the cases where
perturbation method is invalid. In the figures the time and angular
velocities have dimensionless values by employing a time unit $t_c$ defined
as
\bea
t_c = \sqrt{\frac{I^{xx}}{C}} .
\label{unitt}
\eea  
Then, for demonstrating purpose, the integral constant $\Om^z$ is taken to
be 1, the ratio $I^{zz}/I^{xx}$ is taken to be 1.1, and the eigenvalues of
$\bar s^{ij}$ are taken to be $\bar s^{XX} = 0.02, \, \bar s^{YY}=0.01, \,
\bar s^{ZZ} =-0.04$.

\begin{figure}
    \centering
    \includegraphics[width=5.5cm]{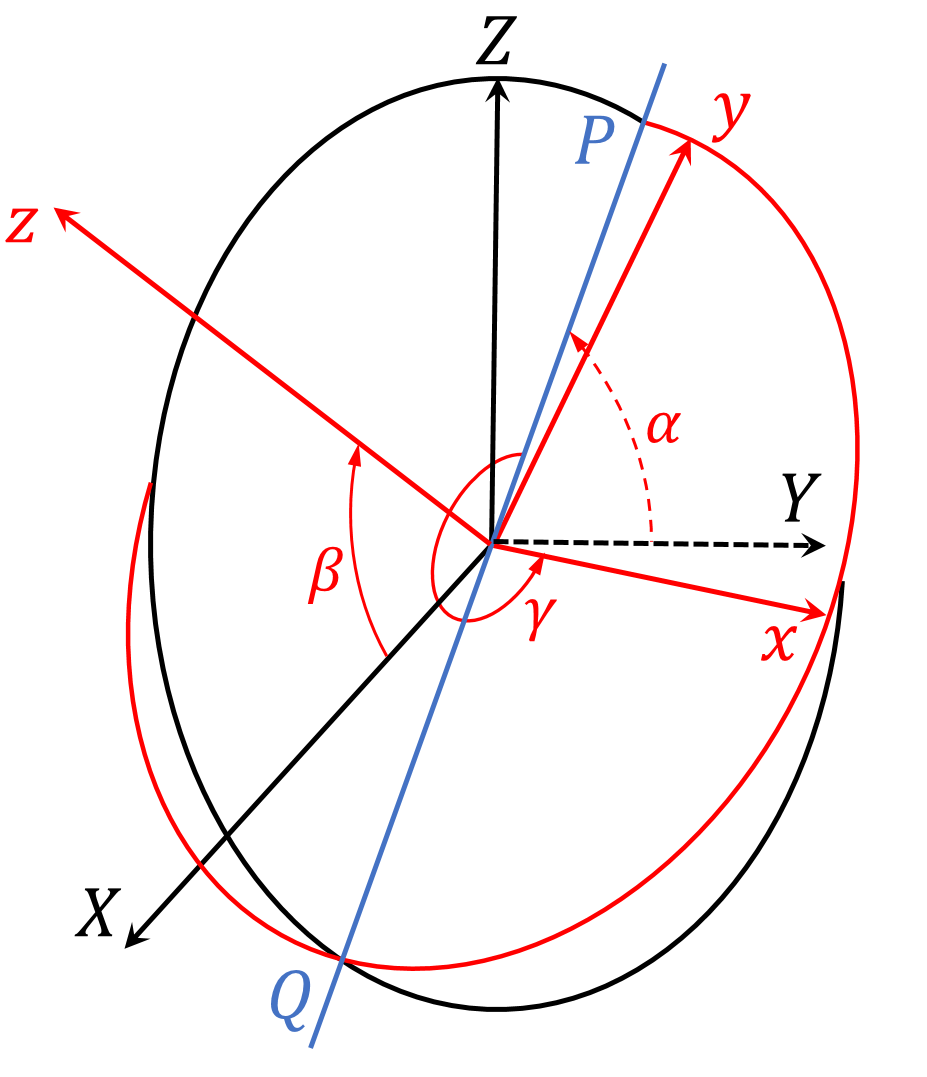}
    \caption{Euler angles $\al, \, \be$, and $\ga$ connecting the
    $Y$-$Z$-$X$ frame and the $x$-$y$-$z$ frame. The line $PQ$ is the
    intersection of the $Y$-$Z$ plane and the $x$-$y$ plane. }
    \label{euler2}
\end{figure}

Before turn to observational consequences, let us address the fact that our
discussion above assumes that the forced precession due to Lorentz
violation is around the $Z$-axis. Depending on the initial values, the
forced precession can be around the other two principal axes of the matrix
$\bar s^{IJ}$. An interesting result from our study of the numerical
solutions is that without loss of generality, if the three different
eigenvalues of $\bar s^{ij}$ satisfy $\bar s^{ZZ} < \bar s^{YY} < \bar
s^{XX}$, then \Reply{the forced precession} around the $Y$-axis is unstable. This
is very similar to the Dzhanibekov effect (also called the tennis racket
theorem) in free rotation~\cite{Louis:1851}. Once this is stated, we point
out that the above discussion equally applies to \Reply{the forced precession} around
the $X$-axis by changing the indices
\bea
Z \rightarrow X, \quad X \rightarrow Y, \quad Y \rightarrow Z ,
\eea
in Eqs.~\rf{consdef} and keeping in mind that $(\al, \be, \ga)$ now refer
to the Euler angles between the $x$-$y$-$z$ frame and the $Y$-$Z$-$X$ frame
(see Fig.~\ref{euler2}).

%%%%%%%%%%%%%%%%%%%%%%%%%%%%%%%%%%%%%%%%%%%%%%%%%%%%%%%%%%%%%%%%%%%%%%%%%%%%%%%%%%%%%%%%%%%
\section{observational consequences}
\label{sec3}

NSs observed as pulsars provide tests against the above
predicted single-precession motion due to Lorentz-violating
gravity~\cite{Nordtvedt:1987, Shao:2013wga, Shao:2014oha,
Shao:2019nso}. In Sec.~\ref{sec3a} we will discuss constraints set on the
coefficients for Lorentz violation by applying the single-precession
solution to two observed solitary
millisecond pulsars~\cite{Shao:2013wga}. On the other hand, the
twofold-precession solution, developed from free precession once
Lorentz-violating gravity presents, has not found its tests as evidences of
free-precessing NSs are yet preliminary (see e.g.
Ref.~\cite{Stairs:2000zz}). \Reply{Based on the hypothesis that NSs might possess tiny rigid deformation,} the modulations on pulsar signals and the
emission of continuous GWs due to free precession have been studied in the
literature to predict observational signatures for searching such
NSs~\cite{Zimmermann:1979ip, 2001MNRAS.324..811J, Gao:2020zcd,Gao:2020dwy}.
Following similar considerations, we investigate in Secs.~\ref{sec3b}
and~\ref{sec3c} observational signatures in pulsar pulses and continuous GW
signals from NSs undergoing the twofold-precession motion.

\subsection{Single-precession motion and constraints on Lorentz violation}
\label{sec3a}

We start with considering the change of the angle between the spin axis
$\hat {\boldsymbol{e}}_z$ and a fixed direction in the $X$-$Y$-$Z$ frame when a star
is in the state of single precession. Calling that angle $\la$ and
describing the fixed direction using the spherical angular coordinates
$(\th_o, \, \ph_o)$ in the $X$-$Y$-$Z$ frame, we have the relation
\bea
\cos\la &=& \sin\th_o \sin\be \sin{(\al - \ph_o)} + \cos\th_o \cos\be ,
\eea 
where $\al$ and $\be$ are Euler angles used in Sec.~\ref{sec2}. Assume that
the changes in $\al$ and $\be$ are much smaller than 1 during a certain
time interval, then the change of $\cos\la$ can be approximated as
\bea
\De \cos\la &\approx& \left( \sin\th_o \cos\be \sin{(\al - \ph_o)} - \cos\th_o \sin\be \right) \De \be 
\nonumber \\
&& + \sin\th_o \sin\be \cos{(\al - \ph_o)} \De \al.
\eea  
For estimations, we use the perturbation solution \rf{ansatz} to
approximate the changes in $\al$ and $\be$ as
\bea
\De \al &\approx&\dot \al \De t = \left( 1 + \frac{A}{a^{(1)} \sin\be_0} \sin{\vp} + \frac{\cos\be_0 \tilde C}{a^{(1)}} \cos{2\al_0} \right) a^{(1)} \De t,
\nonumber \\
\De \be &\approx&\dot \be \De t = \left( \frac{A}{a^{(1)}} \cos{\vp} + \frac{\sin\be_0 \tilde C}{a^{(1)}} \sin{2\al_0} \right) a^{(1)} \De t ,
\eea
where the start of the time interval $\De t$ has been set at $t=0$. 
Because $A,\, \tilde C$, and $a^{(1)}$ are at the same order, $\De \al, \,
\De\be$, and hence $\De \cos\la$ are proportional to $a^{(1)} \De t$ with
factors of order unity. Here we neglect any situation where $(\th_o,\,
\ph_o)$ and $\{A,\, \vp, \, \al_0, \, \be_0\}$ are fine-tuned to vanish the
factor for $\De \cos\la$. Once an observation puts a bound on the
change of $\la$ during a certain time interval, corresponding constraints
on the combination $2\bar s^{ZZ} - \bar s^{XX} - \bar s^{YY}$ is
\bea
\left| 2\bar s^{ZZ} - \bar s^{XX} - \bar s^{YY} \right| \lesssim  \left| \frac{I^{zz} \Om^{z} }{C \De t \cos\be_0} \, \De\cos\la \right| .
\label{est}
\eea

The estimation \rf{est} depends on $\be_0$, which roughly characterizes the
angle between the symmetric axis of the star and the Lorentz-violating
principal axis around which the forced precession happens. Checked with
numerical results, we find that when $\be_0 \lesssim 1.1 $, the estimation
\rf{est} gives correct orders of magnitude for $\left| 2\bar s^{ZZ} - \bar
s^{XX} - \bar s^{YY} \right|$. But as we mentioned in Sec.~\ref{sec2b},
when $\be_0$ approaches $\pi/2$, the estimation \rf{est} fails because the
perturbation method breaks down. A semianalytical relation based on
Eq.~\rf{est} and numerical results is
\bea
\left| 2\bar s^{ZZ} - \bar s^{XX} - \bar s^{YY} \right| \lesssim \et \left| \frac{I^{zz} \Om^{z} }{C \De t} \, \De\cos\la \right| , 
\label{est1}
\eea
where $\et$ can be approximated by $1/\cos\be_0$ for $\be_0 \lesssim 1.1 $
but then only increases to $10$ when $\be_0$ approaches $\pi/2$.

\begin{table*}
    \caption{Observational quantities from PSRs~B1937+21 and
    J1744$-$1134~\cite{Shao:2013wga} to constrain Lorentz violation. The
    pulse profile of PSR~B1937+21 consists of a main pulse and an
    interpulse, both of which put bounds on the change of $\la$ and hence
    constrain Lorentz violation. The pulse width is taken at $50\%$ intense
    level in practice. The last row, conservative bound on $\left| 2\bar s^{ZZ} - \bar s^{XX} - \bar s^{YY} \right|$, is obtained by setting $\et = 10$.  \label{tab2}}
    \def\arraystretch{1.3}
    \begin{tabularx}{\textwidth}{p{5.7cm}p{4.5cm}p{4.5cm}p{3cm}}
      \hline   \hline
        & Main pulse of PSR B1937+21 & Interpulse of PSR B1937+21 & PSR J1744$-$1134 \\
      \hline
      Spin period (ms)  & $1.6$ & $1.6$ & $4.1$ \\ 
      $W$ (deg)         & $8.3$ & $10$ & $13$ \\
      Bound on $\De W$ ($10^{-3}$ deg)     & $-48$ & $53$ & $20$ \\
      $\la$ (deg)  & $100$ & $100$ & $95$ \\
      $\ch$ (deg) & 75 & $105$ & 51 \\
      Bound on $\De \la$ ($10^{-3}$ deg)   & $-19$ & $-5.2$ & 1.6 \\
      Bound on $\left| 2\bar s^{ZZ} - \bar s^{XX} - \bar s^{YY} \right|$  & $10^{-14}\et$ & $10^{-15}\et$ & $10^{-16}\et$ \\
      Conservative bound on $\left| 2\bar s^{ZZ} - \bar s^{XX} - \bar s^{YY} \right|$  & $10^{-13}$ & $10^{-14}$ & $10^{-15}$ \\
      \hline
  \end{tabularx}
\end{table*}

Now we are ready to apply Eq.~\rf{est1} to two solitary pulsars: PSRs~B1937+21 and
J1744$-$1134. Their observed pulse profiles over 15 years were thoroughly
analyzed by \citet{Shao:2013wga}. Here we simply use the conclusion that
the change in the angle $\la$, now being the angle between the spin axis
and the line of sight, is bounded by
\bea
\De \la = \frac{1}{2} \frac{\sin{\frac{W}{2}}}{\cot\la\cos{\frac{W}{2}}+\cot\ch} \, \De W ,
\label{dela}
\eea   
where $W$ is the pulse width, $\De W$ is the change in $W$ during the 15
years, and $\ch$ is the angle between the spin axis of the NS and the
symmetric axis of the pulsar beam which has been assumed to take the shape
of a narrow cone~\cite{Lorimer:2005misc}. The values of those quantities
for each pulsar are listed in Table \ref{tab2}. The derived bounds on $\De
\la$ and hence on the combination $\left|
2\bar s^{ZZ} - \bar s^{XX} - \bar s^{YY} \right|$ are shown in the last three
rows of the table (the last row is the same as the last second row except that $\et$ is set to $10$, which is suggested as the upper limit of $\et$ by our numerical solutions). As an estimation, a uniform density of $\rh_{\rm NS} \sim 10^{15} {\rm g}/{\rm cm}^3$ and a fluid deformation of $\ep_f \sim 10^{-3}$ have been used to get 
\bea
\frac{I^{zz} }{C} \approx \frac{15}{4\pi} \frac{1}{ \rh_{\rm NS} \ep_f} \sim 10^{-5} \, {\rm s}^2,
\eea 
for the NSs.

Under the correspondence \rf{replace}, the constraints in Table~\ref{tab2}
are consistent with those in Ref.~\cite{Shao:2013wga} for the PPN $\al_2$
parameter. Then we do notice that they are 3 to 5 orders of magnitude
better than the {\it global} results in Ref.~\cite{Shao:2014oha} where the
same two solitary pulsars were used but the test was done together with
another 11 binary pulsars in order to obtain global constraints on the
coefficients for Lorentz violation instead of the ``maximal-reach'' ones as
is done here~\cite{Tasson:2019kuw}. This shows that, (i) observations of
solitary pulsars are more sensitive to Lorentz violation than those of
binary pulsars, and (ii) the correlations between different coefficients
for Lorentz violation can severely degrade the
constraints~\cite{Shao:2014oha}.

\subsection{Twofold-precession motion and pulsar signal}
\label{sec3b}

\begin{figure}
    \centering
    \includegraphics[width=8.5cm]{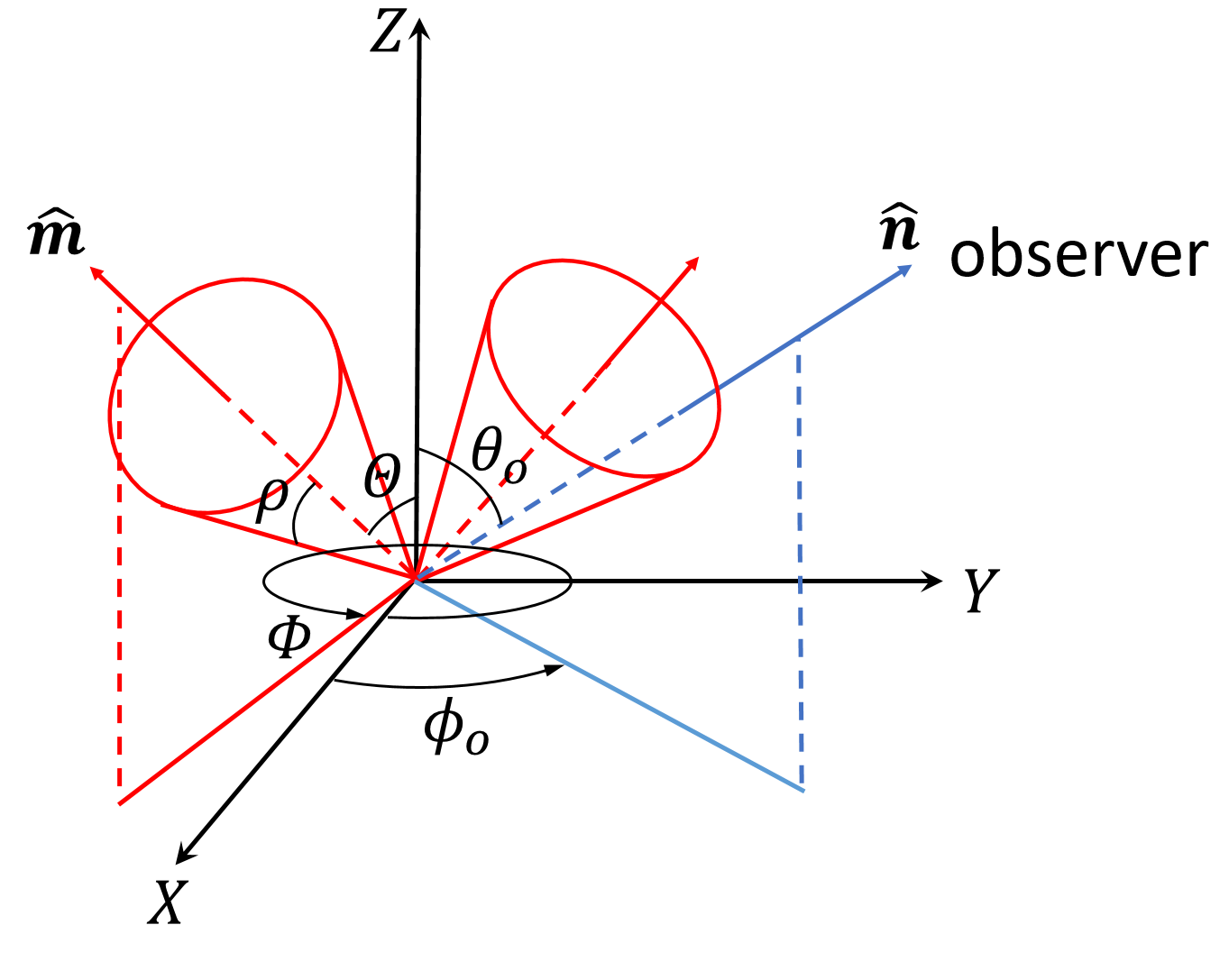}
    \caption{Cone model for pulsar radiation~\cite{1984A&A...132..312G,
    Lorimer:2005misc}. The unit vector along the magnetic dipole moment is
    denoted as $\hat {\boldsymbol{m}}$. It is the symmetric axis of the
    cone. The semi-open angle of the cone is denoted as $\rh$. While the
    star rotates, $\hat {\boldsymbol{m}}$ changes its orientation, and at
    some time, the cone encloses a unit vector $\hat {\boldsymbol{n}}$, so
    the radiation is received by the observer in the direction of $\hat
    {\boldsymbol{n}}$. The colatitude and the azimuth for $\hat
    {\boldsymbol{m}}$ and $\hat {\boldsymbol{n}}$ in the $X$-$Y$-$Z$ frame
    are denoted as $(\Th, \, \Ph)$ and $(\th_o, \, \ph_o)$ respectively. }
    \label{figcone}
\end{figure}

To construct pulsar signals from a NS under the twofold-precession motion,
we adopt the cone model to describe the radiation beam
\cite{1984A&A...132..312G, Lorimer:2005misc}. Figure \ref{figcone}
illustrates a half radiation cone in the $X$-$Y$-$Z$ frame. In principle, the radiation comes from two opposite sides of a NS and is in the shape of a doule cone. But for clarity, we will only track the half cone as shown in Fig.~\ref{figcone} and analyze the signal from it. The signal from the opposite half cone can be obtained simply by reversing the unit vector along the axis of the cone in the following analysis. 

In the cone model, signals are observed when the line of sight
is inside the radiation cone. Mathematically, it means
\bea
\cos\rh < \hat {\boldsymbol{m}} \cdot \hat {\boldsymbol{n}} ,
\label{ineq}
\eea 
where $\rh$ is the semi-open angle of the cone, $\hat {\boldsymbol{m}}$ is
the unit vector along the axis of the cone, and $\hat {\boldsymbol{n}}$ is
the unit vector along the line of sight. The axis of the cone, generally
believed to be aligned with the magnetic dipole moment of the star, is
fixed in the body frame and therefore can be described by two constant
angular coordinates $(\ch, \de)$ as
\bea
\hat {\boldsymbol{m}} = \sin\ch \cos\de \, \hat {\boldsymbol{e}}_{x} + \sin\ch \sin\de \, \hat {\boldsymbol{e}}_{y} + \cos\ch \, \hat {\boldsymbol{e}}_{z} . 
\eea
The direction of the line of sight, if we neglect the proper motion of the
star relative to the observer, is fixed in the $X$-$Y$-$Z$ frame and will be
described by two constant angular coordinates $(\th_o, \ph_o)$ as
\bea
\hat {\boldsymbol{n}} = \sin\th_o \cos\ph_o \, \hat {\boldsymbol{e}}_{X} + \sin\th_o \sin\ph_o \, \hat {\boldsymbol{e}}_{Y} + \cos\th_o \, \hat {\boldsymbol{e}}_{Z} . 
\eea 
Using Eqs.~\rf{trans}, the product of $ \hat {\boldsymbol{m}}$ and $ \hat
{\boldsymbol{n}}$ can be written as a function of time due to the
time-dependent Euler angles,
\bea
\hat {\boldsymbol{m}} \cdot \hat {\boldsymbol{n}} &=&  \sin\th_o \sin\ch \cos{(\al-\ph_o)} \cos{(\ga+\de)} 
\nonumber \\
&& - \sin\th_o \sin\ch \sin{(\al-\ph_o)} \cos\be\sin{(\ga+\de)}  
\nonumber \\
&& + \cos\th_o \cos\ch \cos\be
\nonumber \\
&& + \sin\th_o \cos\ch \sin{(\al-\ph_o)} \sin\be 
\nonumber \\
&&  + \cos\th_o \sin\ch \sin\be\sin{(\ga+\de)} .
\label{pulsemn}
\eea 
The angles $\ph_o$ and $\de$ can be absorbed into the initial values of
$\al$ and $\ga$ in the above expression. For given $\th_o, \, \ch$, and
$\rh$, once the rotation of the star is known, the inequality \rf{ineq} can
be solved to predict the time intervals during which signals are observed.

A quick review of the observational signatures of free-precessing pulsars
is heuristic to explore the twofold-precession case; more details can be
found in e.g. Ref.~\cite{Gao:2020zcd}. For an axisymmetric NS, the
free-precessing solution has
\bea
\dot \al &=& \frac{I^{zz} \Om^z}{I^{xx} \cos\be}, 
\nonumber \\ 
\dot \ga &=& \Om^z - \dot\al \cos\be = - \ep \Om^z , 
\label{angfreqfre}
\eea  
with the angle $\be$ being constant. Therefore, the expression $\hat
{\boldsymbol{m}} \cdot \hat {\boldsymbol{n}}$ as shown in Eq.~\rf{pulsemn}
is simply a sum of four sinusoids of angular frequencies $\dot\al-\dot\ga,
\, \dot\al, \, \dot\al+\dot\ga$, and $\dot\ga$. The analysis is further
simplified by the fact that the deformation $\ep$ is extremely small for
NSs so that the frequency components $\dot\al\pm\dot\ga$ are basically the
same as $\dot\al$, leaving the pulsar period to be $P \approx 2\pi/ \dot
\al$ and modulated with a period of $P_{P} \approx 2\pi/|\dot\ga|$. Here we
assume that the directions of the $z$-axis and the $Z$-axis are properly
chosen so that $\Om^z$ and $\cos\be$ are positive.

An exact analytical expression of the pulsar period $P$ in terms of the
Euler angles was derived in Ref.~\cite{2001MNRAS.324..811J} by calculating
the rate of the azimuthal angle of $\hat {\boldsymbol{m}}$ in the
$X$-$Y$-$Z$ frame. Denote the angular coordinates of $\hat
{\boldsymbol{m}}$ in the $X$-$Y$-$Z$ frame as $(\Th, \Ph)$, then their
relation with $(\ch, \de)$ can be found by transforming the Cartesian
components of $\hat {\boldsymbol{m}}$ under the rotation \rf{rmat}. The
result is
\bea
\cos\Th &=& \cos\be \cos\ch + \sin\be \sin\ch \sin{(\ga+\de)} ,
\nonumber \\
\tan\Ph &=& \tan{(\al + \De\Ph)} ,
\label{tranang}
\eea  
where 
\bea
\tan\De\Phi = \frac{\cos\be \sin{(\ga+\de)} - \sin\be \cot\ch}{\cos{(\ga+\de)}} .
\eea
The pulsar period, defined as the time intervals between $\hat
{\boldsymbol{m}}$ consecutively passing through the plane formed by the
$Z$-axis and the observer, is then
\bea
P \equiv \frac{2\pi}{\dot \Ph} = \frac{2\pi}{\dot\al + \De\dot\Ph} \approx \frac{2\pi}{\dot\al} \left( 1 - \frac{\De\dot\Ph}{\dot\al} \right) ,
\label{pulsep}
\eea
where
\bea
\De\dot\Ph = - \dot\ga \frac{\sin\be\cot\ch\sin{(\ga+\de)} - \cos\be}{\cos^2{(\ga+\de)} + \left( \sin\be \cot\ch - \cos\be\sin{(\ga+\de)} \right)^2} \,.
\eea
Equation \rf{pulsep} not only confirms that the pulsar period $P$ is
approximately generated by the angular frequency $\dot\al$ and that the
modulation on $P$ involves sinusoidal functions of $\ga$, but also
describes the time evolution of $P$ quantitatively, ready to fit
observational data, say, to extract the parameters $\dot \al$, $\dot \ga$,
$\be$, and $\ch$.

Besides the pulsar period, the widths of pulse signals also encode
information on the motion of the star. With the cone model of the radiation
beam, it is defined as the change of the azimuthal angle of $\hat
{\boldsymbol{m}}$ in the $X$-$Y$-$Z$ frame during the time when $\hat
{\boldsymbol{n}}$ is inside the cone, namely
\bea
W \equiv \Ph_2 - \Ph_1 ,
\eea
where $\Ph_1$ and $\Ph_2$ are the azimuthal angle $\Ph$ solved from the
equation of the inequality \rf{ineq} with
\bea
\hat {\boldsymbol{m}} \cdot \hat {\boldsymbol{n}} = \sin\Th\sin\th_0\cos{(\Ph-\ph_o)} + \cos\Th\cos\th_o .
\label{mninertial}
\eea 
Fortunately, because $|\dot\ga|$ is much smaller than $\dot\Ph \approx
\dot\al$, the angle $\Th$, given by the first equation in Eqs.~\rf{tranang}, can
be treated as unchanged during $\Ph$ changing from $\Ph_1$ to $\Ph_2$ so
that the roots $\Ph_1$ and $\Ph_2$ are symmetric about the azimuthal angle
$\ph_o$ of the observer, namely,
\bea
\Ph_1 \approx \ph_o - \frac{W}{2}, \quad \Ph_2 \approx \ph_o + \frac{W}{2},
\eea
and hence, as given in Refs.~\cite{1984A&A...132..312G, Lorimer:2005misc},
setting the expression~\rf{mninertial} equal to $\cos\rh$ leads to an
analytical expression for $W$,
\bea
\cos{\frac{W}{2}} \approx \frac{\cos\rh - \cos\Th \cos\th_o}{\sin\Th \sin\th_o} .
\label{pulsewid}
\eea
When the time dependence of $\Th$ is put back in Eq.~\rf{pulsewid}, it
describes the modulated pulse width at the period $P_P \approx 2\pi/|\dot
\ga|$. By fitting this equation to observational data, the parameters
$\dot\ga, \, \rh$, and $\th_o$ can be extracted.

\begin{figure*}
    \centering
    \includegraphics[width=16cm]{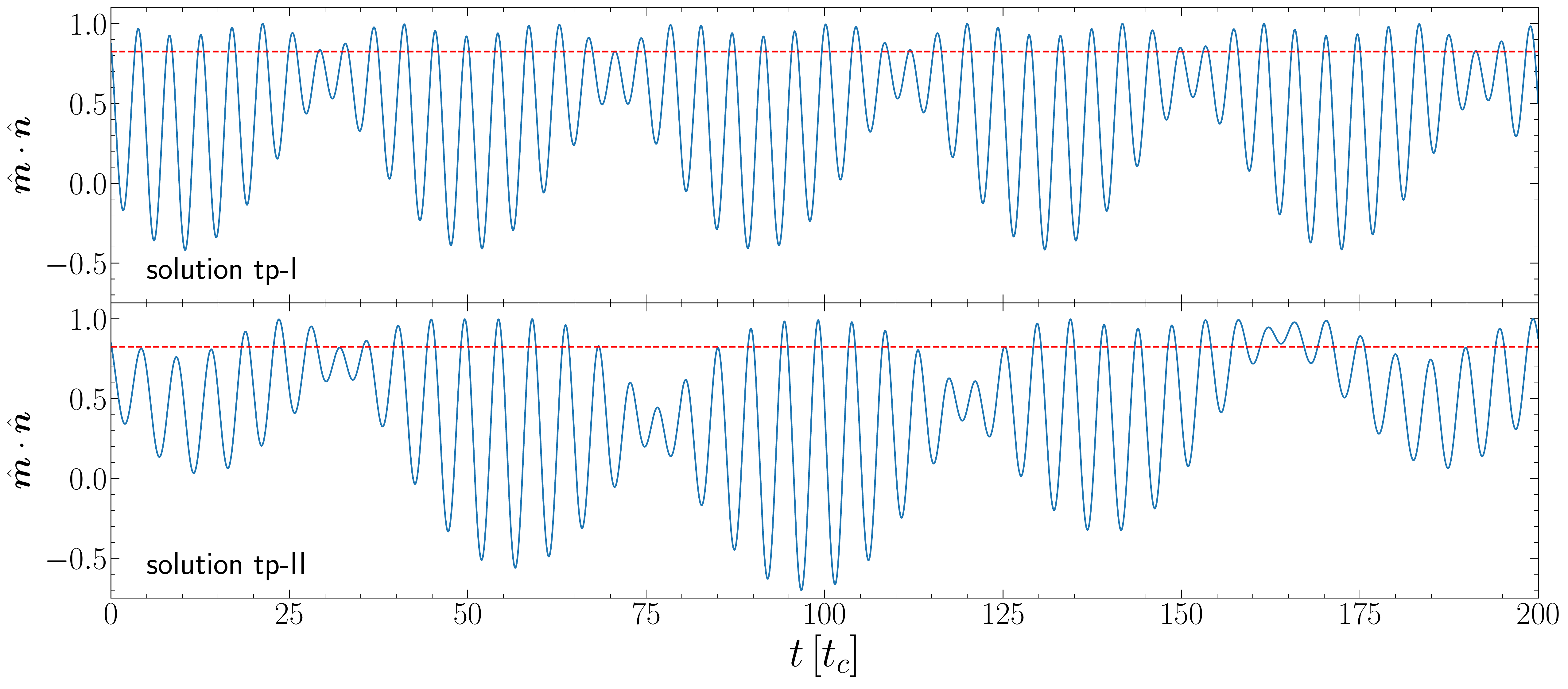}
    \caption{Time evolutions of $\hat {\boldsymbol{m}} \cdot \hat
    {\boldsymbol{n}}$ using the numerical solutions in Figs.~\ref{fig2} and \ref{fig3}. The constant angles $\th_o$ and $\ch$ are taken to be $0.8\,
    {\rm rad}$ and $0.5\, {\rm rad}$ in the plots. The red dashed lines
    represent the cosine value of $\rh = 0.6\, {\rm rad}$, illustrating the
    time segments when $\hat {\boldsymbol{m}} \cdot \hat {\boldsymbol{n}}$
    is above $\cos\rh$ so that the signal is observed under the
    corresponding setup. \Reply{With our choice of $\ep_r = 0.1$ in the numerical solutions, the time unit is $t_c \sim 10^{-3}\,$s if the density of the NS is $\rh_{\rm NS} \sim 10^{15} \, {\rm g}/{\rm cm}^3$.} }
    \label{mnfig1}
\end{figure*}

Let us now turn to the Lorentz-violating twofold-precession case. Similar
characteristics of pulse signals exist. In fact, Eqs.~\rf{pulsep} and
\rf{pulsewid} apply straightforwardly as long as $|\dot\ga| \ll \dot\al$.
This is exactly the regime where the perturbation solution of the twofold
precession works. Neglecting the oscillating terms, $\al^{(1)}$ and
$\be^{(1)}$, the perturbation solution has
\bea
\dot \al &\approx& a = \frac{I^{zz} \Om^z}{I^{xx}\cos\be_0} - a^{(1)} ,
\nonumber \\
\dot\ga &=& \Om^z - \dot\al \cos\be \approx -\ep \Om^z + \cos\be_0 \, a^{(1)} ,
\label{tpconstant}
\eea
where the free-precessing angular frequencies, $\dot \al$ and $\dot \ga$,
are shifted due to Lorentz violation. By fitting Eqs.~\rf{pulsep} and
\rf{pulsewid} to observational data, we will be able to extract the two shifted
frequencies, $\dot\al$ and $\dot\ga$, and the angular parameters $\be_0$,
$\ch$, $\rh$, and $\th_o$.

\begin{figure}
    \centering
    \includegraphics[width=8cm]{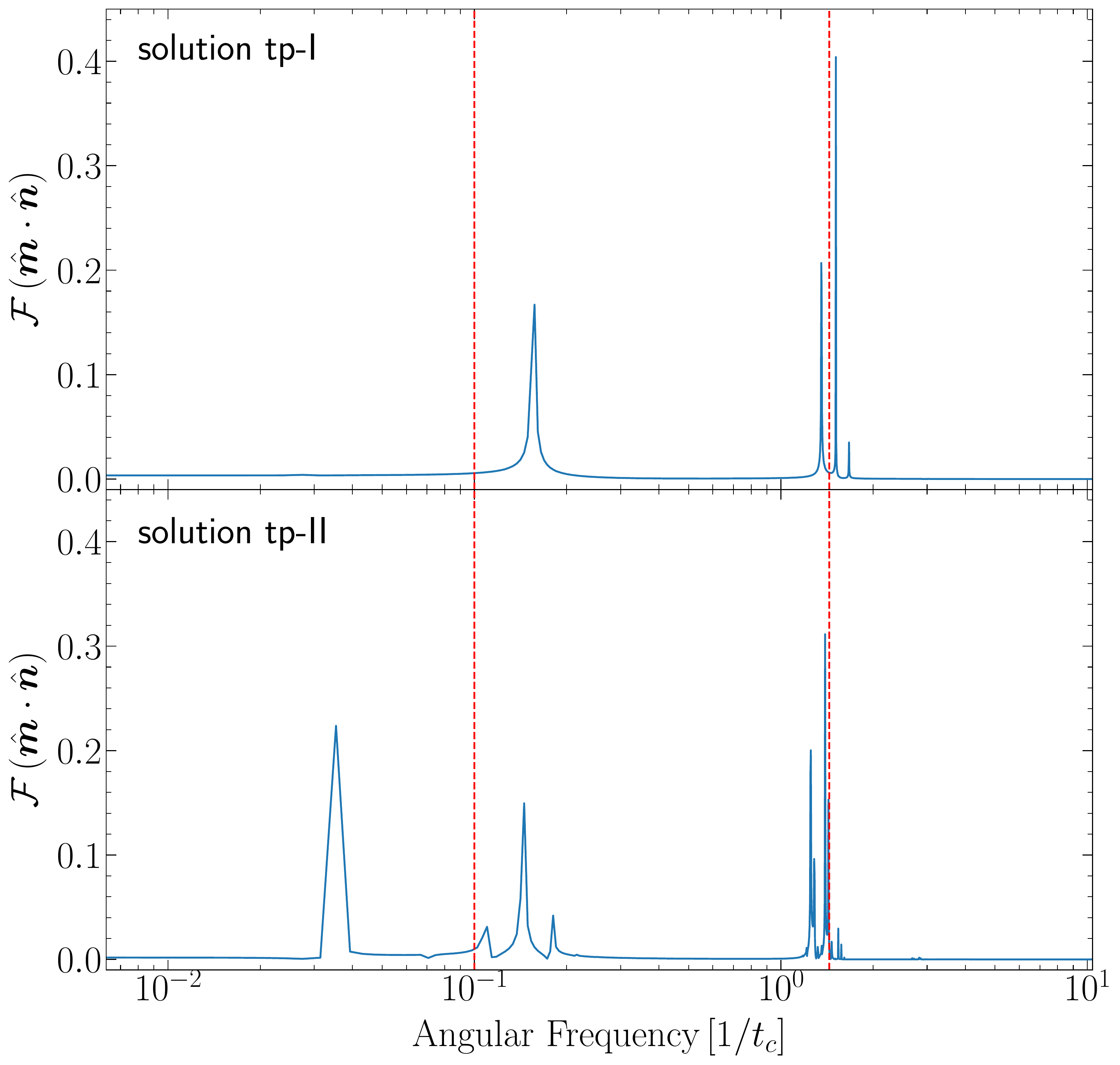}
    \caption{Fourier transformations for $\hat {\boldsymbol{m}} \cdot \hat
    {\boldsymbol{n}}$ in Fig.~\ref{mnfig1}. The dashed red vertical lines mark $\dot \al \approx 1.438$
    and $|\dot \ga| = 0.1$, which are the values from the free-precessing
    equations \rf{angfreqfre} with $I^{zz}/I^{xx} = 1.1$, $\Om^z = 1$, and
    $\be =0.7$. \Reply{As we have taken $\ep_r = 0.1$, the frequency unit is $1/t_c \sim 1000 \, {\rm rad/s}$ if the density of the NS is $\rh_{\rm NS} \sim 10^{15} \, {\rm g}/{\rm cm}^3$.} }
    \label{mnfig2}
\end{figure}

\begin{figure*}
    \centering
    \includegraphics[width=15cm]{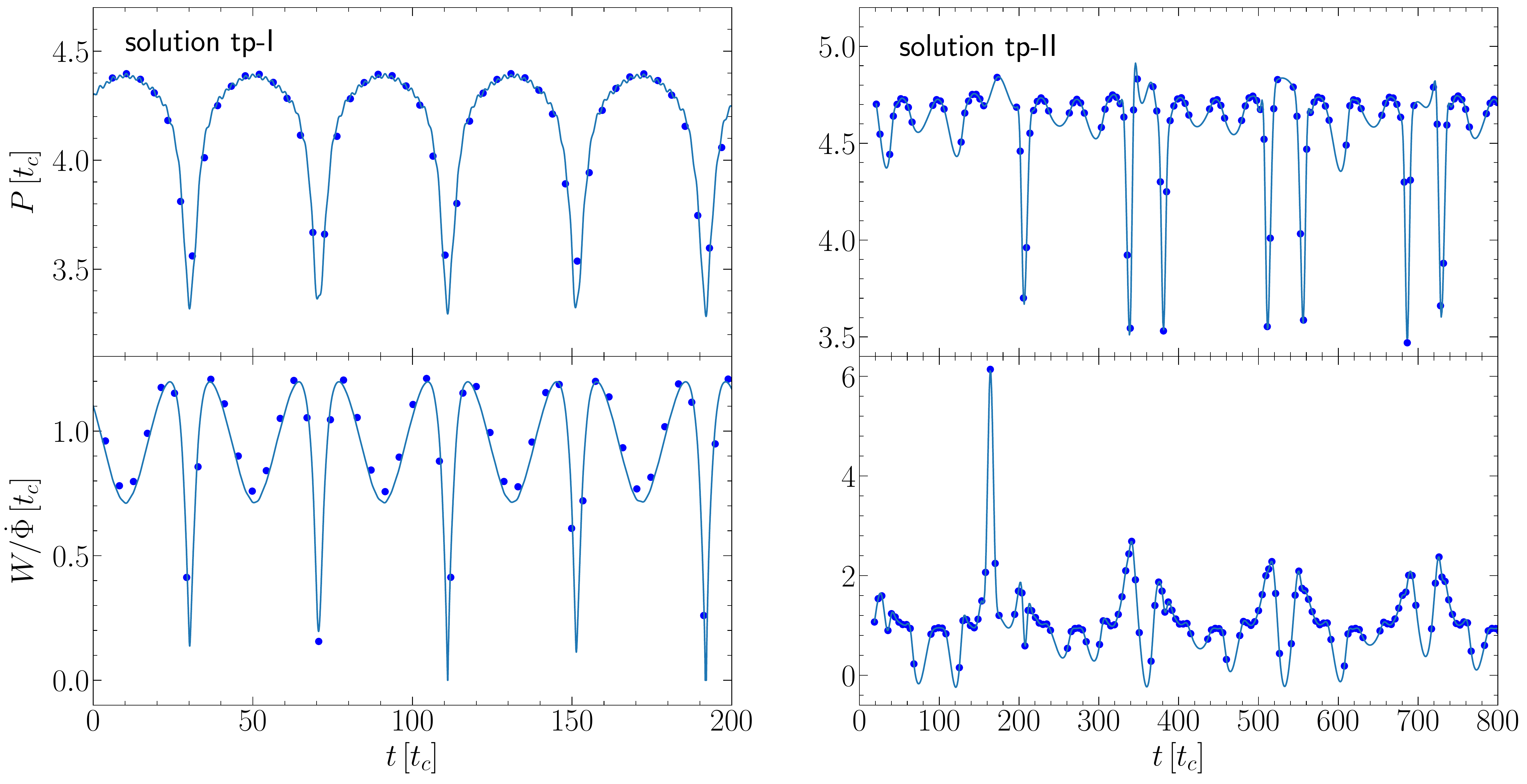}
    \caption{Discrete time series of pulsar period and time segment
    associated to pulse width, solved from $\hat {\boldsymbol{m}} \cdot
    \hat {\boldsymbol{n}} = \cos\rh$ with $\hat {\boldsymbol{m}} \cdot \hat
    {\boldsymbol{n}}$ from Fig.~\ref{mnfig1} and $\rh = 0.6\, {\rm rad}$.
    The left plots are for the solution tp-I where the lines are generated
    by Eqs.~\rf{pulsep} and \rf{pulsewid}. The right plots are for the
    solution tp-II where Eqs.~\rf{pulsep} and \rf{pulsewid} fail to
    approximate the discrete points and the lines are spline interpolations
    of the discrete points.}
    \label{mnfig3}
\end{figure*}

A crucial defect here is that once $\dot\al$ and $\dot\ga$ are extracted,
we are unable to deduce $\Om^z$, $\ep$, and $a^{(1)}$ simultaneously. One
more piece of information on the three quantities is required from the
observational data. It involves details of the pattern of $\hat
{\boldsymbol{m}} \cdot \hat {\boldsymbol{n}}$ that are beyond the
perturbation approximation of $\dot\al$ and $\dot\ga$ in
Eqs.~\rf{tpconstant}, and inevitably invokes numerical solutions of the
twofold-precession motion. Numerical calculations for $\hat {\boldsymbol{m}}
\cdot \hat {\boldsymbol{n}}$ are also necessary for another important
reason. When $\dot\ga$ is not much smaller than $\dot\al$, the perturbation
solution breaks down, and to make matters worse, Eqs.~\rf{pulsep} and
\rf{pulsewid} are also not valid any more, so there are no analytical
templates to fit the pulsar period and the pulse width. In this case, the
inequality~\rf{ineq} needs to be solved numerically to deduce discrete time
series of pulsar period and pulse width to fit observational data. The
rotation of the star, the parameters $\ep$, $\ch$, $\rh$, and $\th_o$, and
the coefficients for Lorentz violation $\bar s^{XX}$, $\bar s^{YY}$, and $\bar
s^{ZZ}$, might be determined via elaborate numerical calculations and
careful treatments. Of course, other measurements, like the polarization
properties of the radiation, might provide extra information and can be
combined to derive a full solution. This lays beyond the scope of the current
work.

In Fig.~\ref{mnfig1}, the time evolutions of $\hat {\boldsymbol{m}}
\cdot \hat {\boldsymbol{n}}$ using the two numerical solutions in
Fig.~\ref{fig2} (solution tp-I) and Fig.~\ref{fig3} (solution tp-II) are displayed, while
Fig.~\ref{mnfig2} depicts the corresponding angular frequency spectra. A
decisive characteristic shows up in the case of the solution that
invalidates the perturbation approach (namely, solution tp-II). That is the
extra smallest frequency component appearing in the lower plot of
Fig.~\ref{mnfig2}. This frequency corresponds to the average angular
velocity at which the angular momentum of the star precesses about the
$Z$-axis. Though no simple approximate expression relates it to the
coefficients for Lorentz violation, numerical calculations indicate that
this angular frequency has the same order of magnitude as the coefficients
for Lorentz violation under the dimensionless parametrization using the
time unit defined in Eq.~\rf{unitt}, \Reply{which for realistic NSs has an estimation} 
\bea
t_c \approx \sqrt{\frac{15}{4\pi} \frac{1}{\rh_{\rm NS} \ep_r}} \sim 1 {\rm s},
\eea
with density $\rh_{\rm NS} \sim 10^{15}\, {\rm g}/{\rm cm}^3$ and rigid
deformation $\ep_r \sim 10^{-8}$ assumed.\footnote{\Reply{We must point out that because we use $I^{zz}/I^{xx} = 1.1$ for illustrative purposes in our numerical examples, meaning $\ep_r = 0.1$, the time unit $t_c$ in the plots is really at the order of $10^{-3}\,$s if the density is $\rh_{\rm NS} \sim 10^{15} \,{\rm g}/{\rm cm}^3$.}} The extra frequency component does not
show up in the case of solution tp-I. Because in that case, the angular
momentum of the star is almost aligned with the $Z$-axis, so the precession
of the angular momentum around the $Z$-axis caused by Lorentz violation
degenerates with the free precession, generating a total precession rate
$\dot\al$ approximately given by the first equation in Eqs.~\rf{tpconstant}
where the Lorentz-violating contribution $a^{(1)}$ is unable to be decoupled.

In Fig.~\ref{mnfig3}, discrete time series of pulsar period and of time
segment corresponding to pulse width calculated by solving the inequality
\rf{ineq} using the numerical templates of $\hat {\boldsymbol{m}} \cdot
\hat {\boldsymbol{n}}$ from Fig.~\ref{mnfig1} are plotted. For the case of
the solution tp-I (left panels of Fig.~\ref{mnfig3}), the smooth curves
that approximate the discrete points are generated using Eqs.~\rf{pulsep}
and \rf{pulsewid}. They fit very well. For the case of the solution tp-II
(right panels of Fig.~\ref{mnfig3}), Eqs.~\rf{pulsep} and \rf{pulsewid}
fail to approximate the discrete series, so we have to interpolate the
points to reveal the trends of them. Note that in both cases there is a period about $40\, t_c$ originated from the angular frequency between $0.1\, t_c^{-1}$
and $0.2\,t_c^{-1}$ in Fig.~\ref{mnfig2}, while in the right panels a
period about $180\, t_c$ shows up---more perceivable in the pulsar-period plot than in the pulse-width plot---reflecting the characteristic angular
frequency between $0.03 \, t_c^{-1}$ and $0.04 \, t_c^{-1}$ in the lower
panel of Fig.~\ref{mnfig2}.

\subsection{Twofold-precession motion and continuous GWs}
\label{sec3c}

The quadrupole GW radiated by a freely precessing rigid body was calculated
in Refs.~\cite{Zimmermann:1979ip, Zimmermann:1980ba}. To investigate the
quadrupole radiation by a rigid body undergoing the twofold-precession
motion due to Lorentz violation, we generalize Zimmermann's
calculation~\cite{Zimmermann:1980ba} to any rotating rigid body with
torques.

\begin{figure*}
    \centering
    \includegraphics[width=17cm]{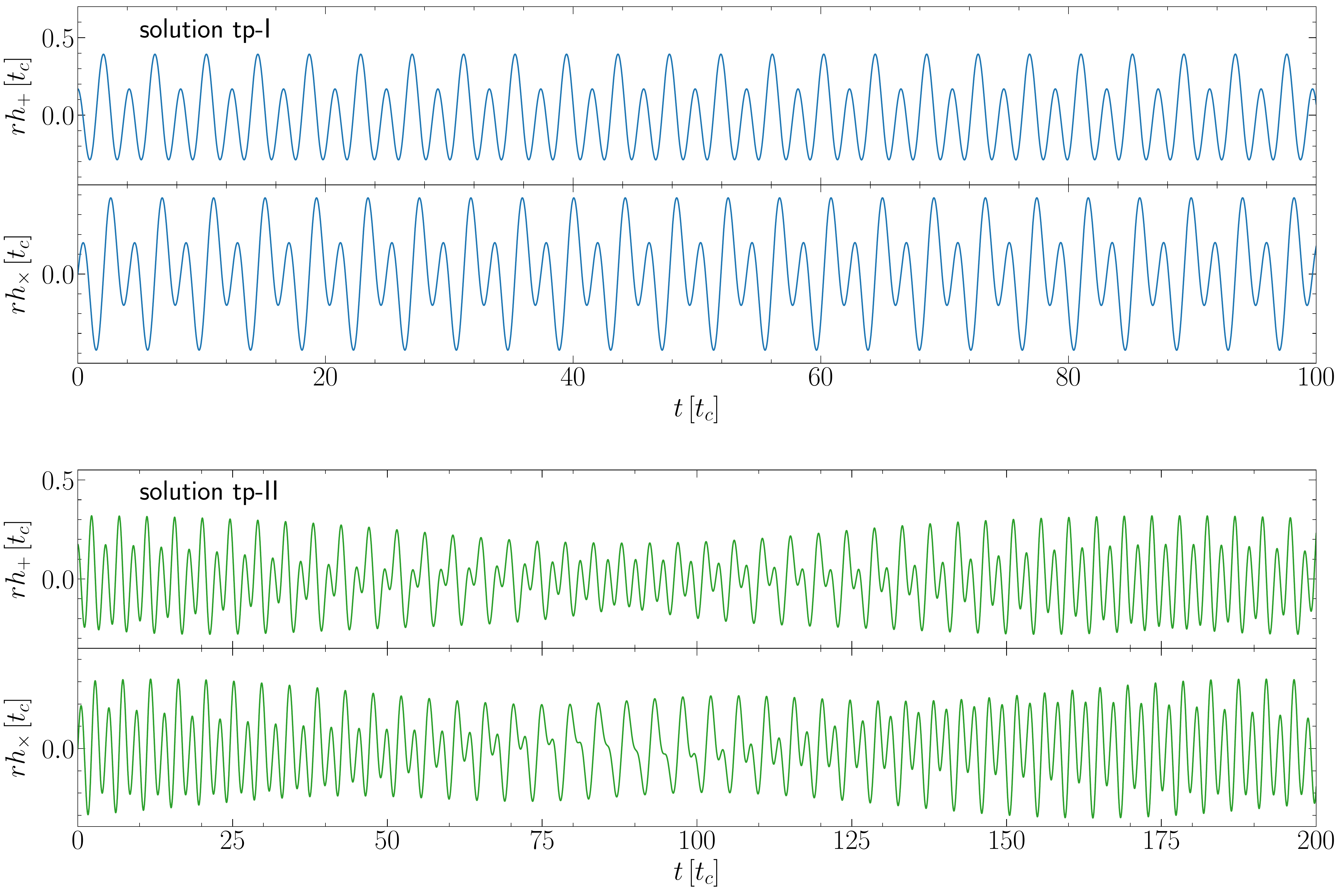}
    \caption{Continuous GWs from NSs subjected to the solutions
    tp-I and tp-II. The angular coordinates of the
    observer are taken to be $\th_o = 0.8\, {\rm rad}$ and $\ph_o = 0$ in
    the plots. Time and distance are parametrized using the time unit $t_c$
    defined in Eq.~\rf{unitt}, \Reply{which is at the order of $10^{-3}\,$s with our choice of $\ep_r = 0.1$ in the numerical solutions and $\rh_{\rm NS} \sim 10^{15} \, {\rm g}/{\rm cm}^3$ assumed for the density of the NS.} }
    \label{gwfig1}
\end{figure*}

The metric perturbation in the $X$-$Y$-$Z$ frame is given by the quadrupole
formula
\bea
h^{IJ} = - \frac{2}{r} \ddot I^{IJ} ,
\eea
where $r$ is the distance from the star to the observer, and the double
dots denote the second time derivative. The time dependence of the
inertial-frame components of the moment of inertia tensor originates in the
rotation matrix via
\bea
I^{IJ} = R^{iI} R^{jJ} I^{ij},
\eea 
while the body-frame components of the moment of inertia tensor are
constant. The derivatives of $R^{iI}$ can be derived by noticing that the
body-frame axes undergo an infinitesimal rotation
\bea
\hat {\boldsymbol{e}}_i \rightarrow \hat {\boldsymbol{e}}_i + dt \, \boldsymbol{\Om} \times \hat {\boldsymbol{e}}_i ,
\eea
at the instant when the body rotates with an angular velocity
$\boldsymbol{\Om}$. The change of $R^{iI}$ is therefore $dt \, \left(
\boldsymbol{\Om} \times \hat {\boldsymbol{e}}_i \right) \cdot \hat
{\boldsymbol{e}}_I $, giving
\bea
\frac{ dR^{iI} }{dt} = \left( \boldsymbol{\Om} \times \hat {\boldsymbol{e}}_i \right) \cdot \hat {\boldsymbol{e}}_I = \ep^{IJK} \Om^J R^{iK} ,
\label{1stdrmat}
\eea 
where $\ep^{IJK}$ is the Levi-Civita symbol. Equation \rf{1stdrmat} can be
used to calculate $\dot I^{IJ}$. When combined with the equation of motion
\bea
\dot I^{IJ} \Om^J + I^{IJ} \dot\Om^J = \Ga^I ,
\eea
$\boldsymbol{\dot\Om}$ can be solved in terms of $\boldsymbol{\Om}$ and the
torque $\boldsymbol{\Ga}$. Then, the demanded second derivative $\ddot
I^{IJ}$ can be found to take the form
\bea
\ddot I^{IJ} = R^{iI} R^{jJ} A^{ij} ,
\eea
with the help of the derivative of Eq.~\rf{1stdrmat}. The body-frame tensor
$A^{ij}$ is lengthy in general, but much simplified when the body frame
diagonalizes the moment of inertia tensor, which is the case in our setup. The components then read
\bea
A^{xx} &=& 2\left( \De_2 \left(\Om^y\right)^2 - \De_3 \left(\Om^z\right)^2 \right) ,
\nonumber \\
A^{yy} &=& 2\left( \De_3 \left(\Om^z\right)^2 - \De_1 \left(\Om^x\right)^2 \right) ,
\nonumber \\
A^{zz} &=& 2\left( \De_1 \left(\Om^x\right)^2 - \De_2 \left(\Om^y\right)^2 \right) ,
\nonumber \\
A^{xy} &=& \left( \frac{ \left(\De_3 \right)^2}{I^{zz}} + \De_1 - \De_2 \right) \Om^x \Om^y + \frac{\De_3}{I^{zz}} \Ga^z,
\nonumber \\
A^{xz} &=& \left( \frac{ \left(\De_2 \right)^2}{I^{yy}} + \De_3 - \De_1 \right) \Om^x \Om^z + \frac{\De_2}{I^{yy}} \Ga^y ,
\nonumber \\
A^{yz} &=& \left( \frac{ \left(\De_1 \right)^2}{I^{xx}} + \De_2 - \De_3 \right) \Om^y \Om^z + \frac{\De_1}{I^{xx}} \Ga^x ,
\label{gwa}
\eea
where $( \Om^x, \, \Om^y, \, \Om^z)$ are the body-frame velocity components
given in terms of the Euler angles and their derivatives in
Eqs.~\rf{angvel1}, and the body-frame torque components $(\Ga^x, \, \Ga^y,
\, \Ga^z)$ can be calculated from the anisotropic potential $\de U$ via
\bea
\Ga^x &=& -\frac{\sin\ga}{\sin\be} \, \prt_\al \de U - \cos\ga \, \prt_\be \de U + \cot\be\sin\ga \, \prt_\ga \de U, 
\nonumber \\
\Ga^y &=& -\frac{\cos\ga}{\sin\be} \, \prt_\al \de U + \sin\ga \, \prt_\be \de U + \cot\be\cos\ga \, \prt_\ga \de U, 
\nonumber \\
\Ga^z &=& - \prt_\ga \de U .
\eea
The symbols $\De_1,\, \De_2$, and $\De_3$ follow the definition of
\citet{Zimmermann:1980ba},
\bea
\De_1 = I^{yy}-I^{zz}, \quad \De_2 = I^{zz}-I^{xx}, \quad \De_3 = I^{xx}-I^{yy}.
\eea
It is plain that Eqs.~\rf{gwa} go back to the result of
\citet{Zimmermann:1980ba} when $\boldsymbol{\Ga} = 0$.

\begin{figure}
    \centering
    \includegraphics[width=8cm]{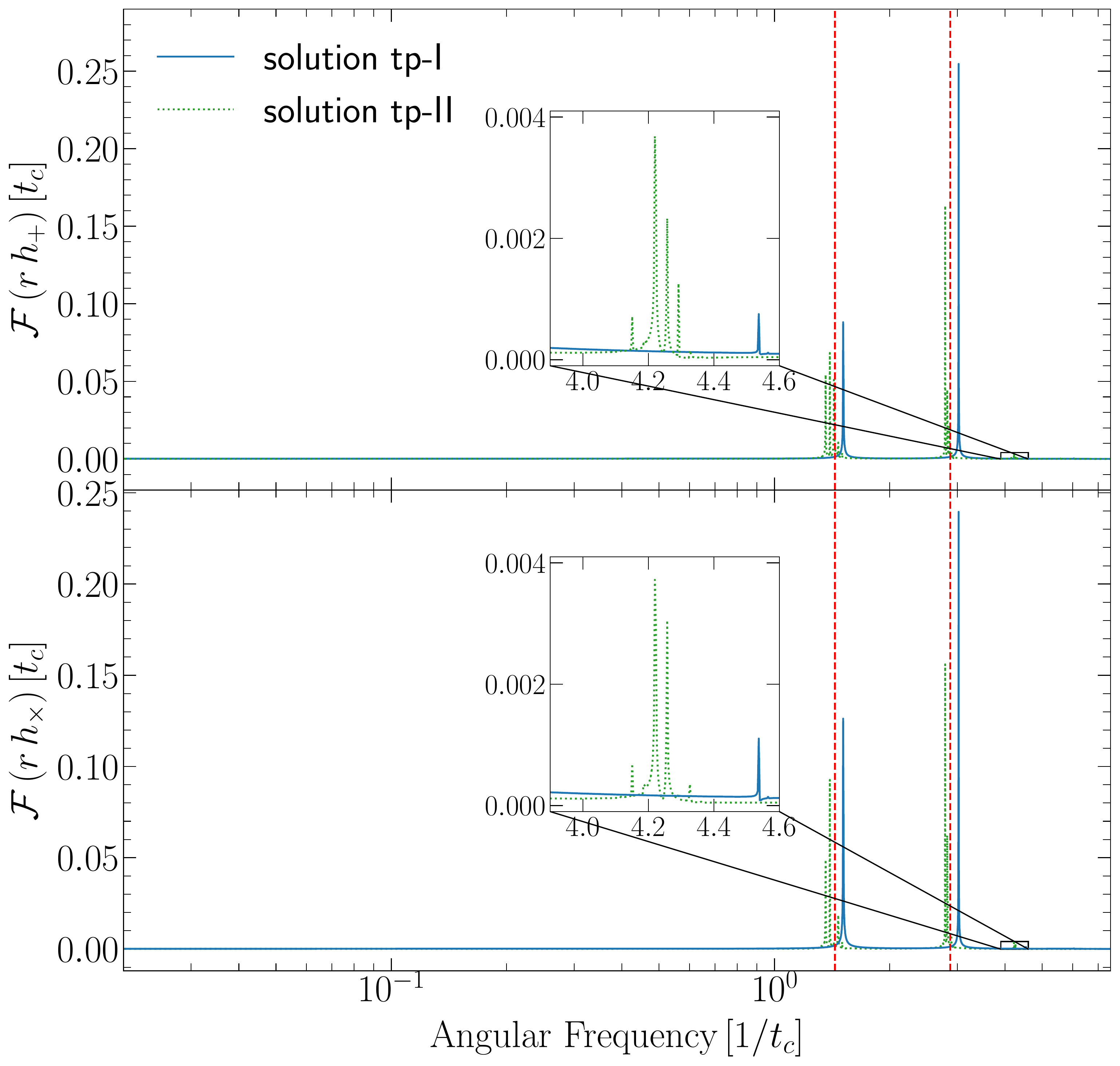}
    \caption{Fourier transformations for the continuous GWs shown in
    Fig.~\ref{gwfig1}. The dashed red vertical lines mark $\dot \al \approx
    1.438$ and $2\dot \al \approx 2.876 $, which are the angular
    frequencies of the continuous GWs from a free-precessing NS with
    $I^{zz}/I^{xx} = 1.1, \, \Om^z = 1$, and $\be =0.7$. \Reply{As we have taken $\ep_r = 0.1$, the frequency unit is $1/t_c \sim 1000 \, {\rm rad/s}$ if the density of the NS is $\rh_{\rm NS} \sim 10^{15} \, {\rm g}/{\rm cm}^3$.} }
    \label{gwfig2}
\end{figure}

As we restrict ourselves to spheroids, the tensor $A^{ij}$ is further
simplified by $I^{xx} = I^{yy}$ and $\Ga^z = 0$. In the absence of Lorentz
violation, the star precesses freely, and the tensor $\ddot I^{IJ}$ has a
simple form,
\bea
\ddot I^{IJ} &=& \frac{ 1}{2}\De_2 \, \dot\al^2 \sin{2\be}
\nonumber \\
&& \times \begin{pmatrix}
2\cos{2\al} \tan\be & 2\sin{2\al} \tan\be & -\sin\al \\
2\sin{2\al} \tan\be & -2\cos{2\al} \tan\be & \cos\al \\
-\sin\al & \cos\al & 0
\end{pmatrix} ,
\label{gwwf}
\eea
which tells that $h^{IJ}$ is a spherical wave with exactly two
frequency components, namely $\dot\al$ and $2\dot\al$. When Lorentz
violation presents and the star is subjected to the twofold-precession
motion, the tensor $\ddot I^{IJ}$ becomes complicated as $\dot\be$ no
longer vanishes. But if we only consider motions given by the perturbation
solution, we expect Eq.~\rf{gwwf} to be a fair approximation and that the
continuous GW still has two frequency components, $\dot\al$ and $2\dot\al$,
with $\dot\al$ being the Lorentz-violating shifted angular frequency given
by the first equation in Eqs.~\rf{tpconstant}.

For an observer with colatitude $\th_o$ and azimuth $\ph_o$ in the
$X$-$Y$-$Z$ frame, the
basis to decompose the two physical degrees of freedom for the continuous
GW can be taken as
\bea
\boldsymbol{e}_{+} \equiv \hat {\boldsymbol{\th}}_o \otimes \hat {\boldsymbol{\th}}_o  - \hat {\boldsymbol{\ph}}_o \otimes \hat {\boldsymbol{\ph}}_o , \quad \boldsymbol{e}_{\times} \equiv \hat {\boldsymbol{\th}}_o \otimes \hat {\boldsymbol{\ph}}_o  + \hat {\boldsymbol{\ph}}_o \otimes \hat {\boldsymbol{\th}}_o ,
\eea
where 
\bea
\hat {\boldsymbol{\th}}_o &=& \cos\th_o \cos\ph_o \, \hat{\boldsymbol{e}}_X + \cos\th_o \sin\ph_o \, \hat{\boldsymbol{e}}_Y - \sin\th_o \hat{\boldsymbol{e}}_Z,
\nonumber \\
\hat {\boldsymbol{\ph}}_o &=& -\sin\ph_o \, \hat{\boldsymbol{e}}_X + \cos\ph_o \, \hat{\boldsymbol{e}}_Y  .
\label{thphinertial}
\eea
Then the ``+'' and the ``$\times$'' components of the continuous GW
are
\bea
h_{+} &=& \frac{1}{2} \left( \hat {\th}_o^I \, \hat \th_o^J  - \hat {\ph}_o^I \, \hat {\ph}_o^J \right) h^{IJ} = - \frac{1}{r} \left( \hat {\th}_o^I \, \hat \th_o^J  - \hat {\ph}_o^I \, \hat {\ph}_o^J \right) \ddot I^{IJ}, 
\nonumber \\
h_{\times} &=& \frac{1}{2} \left( \hat {\th}_o^I \, \hat \ph_o^J  + \hat {\ph}_o^I \, \hat {\th}_o^J \right) h^{IJ} = - \frac{1}{r} \left( \hat {\th}_o^I \, \hat \ph_o^J  + \hat {\ph}_o^I \, \hat {\th}_o^J \right) \ddot I^{IJ},
\label{hpluscross}
\eea
where $\hat {\th}_o^I$ and $\hat {\ph}_o^I$ are the components of $\hat
{\boldsymbol{\th}}_o$ and $\hat {\boldsymbol{\ph}}_o$ in the $X$-$Y$-$Z$
frame as shown in Eqs.~\rf{thphinertial}.

When continuous GWs from NSs are detected, Eqs.~\rf{hpluscross} can be used
as template waveforms to match observational data. Similar to the analysis
of pulsar signal, to extract the coefficients for Lorentz violation, $\bar
s^{XX}$, $\bar s^{YY}$, $\bar s^{ZZ}$, and the NS parameters, $\ep$,
$\Om^z$, $\th_o$, $r$, from observational data, elaborate numerical
calculations are required. Using the two numerical solutions in
Figs.~\ref{fig2} and \ref{fig3} to calculate $\ddot I^{IJ}$, we plot the
corresponding waveforms in Fig.~\ref{gwfig1} and their Fourier
transformations in Fig.~\ref{gwfig2}. Tiny high frequency components in the
spectra are found. They are featured in the subplots of Fig.~\ref{gwfig2}
and turn out to be exactly three times of the fundamental frequencies,
which are represented by the peaks near $1.438 \,t_c^{-1}$ and can be
regarded as the average values of $\dot \al$ for the twofold-precession
motions. Frequencies higher than the third harmonic exist, but their
amplitudes are even much smaller. We point out that unlike the pulsar
signal of solution tp-II, there is no low frequency at the order of the
coefficients for Lorentz violation in the continuous GW spectra. However, the
advantage of continuous GW signal is that both solution tp-I and solution
tp-II exhibit the third harmonic, characterizing the deviation of the waveform
from that of a free-precessing NS, though in the case of solution tp-I, the
amplitude of the third harmonic is further suppressed. These Lorentz-violating
features are worth looking for and may reveal new physics beyond the
current understanding.

%%%%%%%%%%%%%%%%%%%%%%%%%%%%%%%%%%%%%%%%%%%%%%%%%%%%%%%%%%%%%%%%%
\section{conclusion}
\label{sec4}

Lorentz violation modifies the Newtonian potential by an anisotropic
correction, generating a torque on spheroidal stars that forces the
otherwise conserved angular momentum of the star to precess around a
preferred direction defined by the coefficients for Lorentz violation. To solve
the motion rigorously, we first calculate the anisotropic gravitational
self-energy of the star caused by Lorentz violation in the minimal
gravitational SME in Sec.~\ref{sec2a}. The result is proved to be
equivalent to that of \citet{Nordtvedt:1987} when the star is treated as a
stationary spinning fluid star in equilibrium so the tensor virial relation
holds. Discrepancy occurs when the star possesses a rigid deformation which
invalidates the tensor virial relation. Besides stationary spin, free
precession is also a solution of motion for such stars in the absence of
torques.

Then in Sec.~\ref{sec2b}, \Reply{the forced precession} caused by the
Lorentz-violating torque on stationary spinning stars and on
free-precessing stars is explicitly calculated using perturbation method
to solve the rotational equations of motion. Interestingly, we find that
the direction of the forced precession on stationary spinning stars is
opposite to the direction of the forced precession on free-precessing stars
as shown in Eq.~\rf{precangvel}. Numerical solutions are explored to check
the validity of the perturbation approach. Initial values for which the
perturbation approach fails are identified. The study of the solutions is
finished by clarifying that the preferred directions around which \Reply{the
forced precession} happens are exactly the eigenvectors of the matrix $\bar
s^{ij}$. An interesting result is that when the matrix $\bar s^{ij}$
acquires 3 different eigenvalues, the forced precession is unstable if it
is around the eigenvector corresponding to the middle eigenvalue. This is
similar to the well-known Dzhanibekov effect (or the tennis racket
theorem).

After the solutions of motion are studied thoroughly, we apply them to
explore observational consequences for NSs in Sec.~\ref{sec3}.
\Reply{Section \ref{sec3a} treats the two solitary pulsars in Ref.~\cite{Shao:2013wga} as stationary
spinning spheroids at zeroth order, and sets bounds on the coefficients for Lorentz violation by attributing any possible tiny alteration of the NSs'
orientations hidden in observational uncertainties to the forced precession
due to Lorentz violation.} With the connection
between the SME coefficients $\bar s^{ij}$ and the PPN coefficient $\al_2$
shown in Eq.~\rf{replace}, the constraints obtained here are consistent
with the ones for the $\al_2$ coefficient in Ref.~\cite{Shao:2013wga}. However, our ``maximal-reach'' constraints obtained from the
two solitary pulsars are 3 to 5 orders of magnitude better than those in
Ref.~\cite{Shao:2014oha}. The exactly same two solitary pulsars are used
there but together with another 11 binary pulsars for a global analysis.
Notice that here we are using a different coordinate frame from that of
Ref.~\cite{Shao:2014oha} so a plain comparison is only heuristic.
Nevertheless, this suggests that the orbital motions of the binary pulsars
are less sensitive to Lorentz violation than the rotational motions of the
solitary pulsars. Therefore, we urge observers to analyze the stability of
pulsar spins---possibly via the pulse profile stability as it was done by
\citet{Shao:2013wga}---with more suitable systems, and put tighter bounds
on the coefficients for Lorentz violation.

In Sec.~\ref{sec3b} and Sec.~\ref{sec3c}, pulsar signals and continuous GWs
from Lorentz-violating-affected free-precessing NSs are investigated. When
the angular momentum of the star is close to the preferred direction around
which the forced precession happens, the spectra of the signals are very
much like those from a free-precessing NS. The forced precession does
shift frequencies in the spectra, but the contribution from it is
practically unable to be decoupled from the free-precessing frequency
components. When the angular momentum of the star makes a relatively large
angle to the preferred direction around which the forced precession
happens, decisive characteristics that are worth
searching for show up in the spectra. For pulsar signals, the signature is an extremely low frequency
component as shown in the lower plot of Fig.~\ref{mnfig2}. The observables,
namely the pulsar period and the pulse width, are not only modulated by the
average rate of the Euler angle $\ga$, but also modulated by this frequency
(right plots in Fig.~\ref{mnfig3}). The polarization characteristics of
pulsar pulses are interesting to investigate in future studies, and they could provide more information on the rotation of the NS and its radiation properties. For GWs,
the signature is the third harmonic shown in the subplots of
Fig.~\ref{gwfig2}. Free-precessing spheroidal NSs emit continuous GWs only
at the first and the second harmonics of the rate of the Euler angle $\al$.
But once affected by Lorentz violation, the continuous GW emitted by the
star is no longer a simple sum of sinusoidal functions. The waveform
generally involves all harmonics of the average rate of the Euler angle
$\al$, with the amplitudes decrease rapidly after the first two of them.

The characteristic frequencies in the spectra of the signals are
qualitative supports for Lorentz violation if observed. To extract
quantitative values of the coefficients for Lorentz violation, as well as
the physical parameters of the NS, numerical calculations are necessary to
fit observational data. Tentative candidates of free-precessing NSs are
proposed in pulsar observations \cite{Stairs:2000zz, Shabanova:2001ud},
while searches of continuous GWs have not yet confidently identified any
positive detections~\cite{Pisarski:2019vxw, Covas:2020nwy,
Dergachev:2020fli, Papa:2020vfz, Steltner:2020hfd, Zhang:2020rph}. Our study supplies necessary templates for potential new tests of Lorentz violation that take
advantage of the two state-of-the-art observation channels in the era of
multimessenger astronomy. Once free-precessing NSs are unambiguously
identified, the new tests using their modulated pulsar signals and
continuous GWs are bound to enrich our fundamental knowledge on Lorentz
symmetry of our spacetime.

%%%%%%%%%%%%%%%%%%%%%%%%%%%%%%%%%%%%%%%%%%%%%%%%%%%%%%%%%%%%%%%%%%%%%%%%%%%
\begin{acknowledgments}
We thank Quentin G.\ Bailey and V.\ Alan Kosteleck\'y for discussions, \Reply{and anonymous referees for comments}. This work was supported
by \Reply{the National SKA Program of China (2020SKA0120300),} the National Natural Science Foundation of China (11975027, 11991053,
11721303), the Young Elite Scientists Sponsorship Program by the China
Association for Science and Technology (2018QNRC001), the Max Planck
Partner Group Program funded by the Max Planck Society, and the
High-Performance Computing Platform of Peking University. It was partially
supported by the Strategic Priority Research Program of the Chinese Academy
of Sciences through the Grant No. XDB23010200. R.X. is supported by the
Boya Postdoctoral Fellowship at Peking University.
\end{acknowledgments}

%---------------------------------------------------------------------
\appendix
\section{The constant $C$ for uniform spheroids}
\label{app1}

Using the results on Maclaurin spheroids summarized in
Refs.~\cite{1962ApJ...136.1037C, Poisson:2014}, we can explicitly write
down the constant $C$ in terms of the eccentricity defined as
\bea
e \equiv \sqrt{1 - \left( \frac{a_3}{a_1} \right)^2} , 
\label{ecc}
\eea
for a spheroid \rf{spheroid} with uniform density. We start with the
Newtonian potential \rf{newp} inside an ellipsoid of uniform density,
\bea
\Phi = - \pi \rh \left( A_0 - A_1 x^2 - A_2 y^2 - A_3 z^2 \right) .
\eea
The constants $A_0$, $A_1$, $A_2$, and $A_3$ happen to have closed-form
results for spheroids,
\bea
A_0 &=& 2a_1^2 \sqrt{1 - e^2} \frac{\arcsin{e} }{e} = 2a_1^2 \left( 1 - \frac{1}{3} e^2 + O(e^4) \right), 
\nonumber \\
A_1 &=& A_2 =\frac{\sqrt{1-e^2}}{e^2} \left( \frac{\arcsin{e}}{e} - \sqrt{1-e^2} \right)
\nonumber \\
&&  \quad\  =\frac{2}{3} \left( 1 - \frac{1}{5} e^2 + O(e^2) \right) ,
\nonumber \\
A_3 &=& \frac{2}{e^2} \left( 1 - \sqrt{1-e^2} \frac{\arcsin{e}}{e} \right) =\frac{2}{3} \left( 1 + \frac{2}{5} e^2 + O(e^2) \right) .
\eea
It then follows that the constant $C$ is
\bea
C &=& \frac{1}{2} \left( U^{xx}-U^{zz} \right) = \frac{4\pi^2}{15} \rh^2 a_1^2 a_3 \left( a_1^2 A_1 - a_3^2 A_3 \right)
\nonumber \\
&=& \frac{16\pi^2}{225} \rh^2 a_1^4 a_3 \left( e^2 + O(e^4) \right) .
\eea

We point out that the equilibrium condition
\bea
\frac{p}{\rh} = -\Phi + \frac{1}{2} {|\boldsymbol{\Om}}|^2 \left( x^2 + y^2 \right) + {\rm const.}\,, 
\eea
on the surface where $p = 0$ implies
\bea
|{\boldsymbol{\Om}}|^2 = 2\pi \rh \left( A_1 - \frac{a_3^2}{a_1^2} A_3\right) .
\eea   
As the angular velocity ${\boldsymbol{\Om}}$ is along the $z$-axis, we directly see
\bea
T_{\rm rot} = \frac{1}{2} I^{zz} |{\boldsymbol{\Om}}|^2 = 2 C ,
\label{estc}
\eea 
where $I^{zz} = 8\pi/15 \, \rh a_1^4 a_3$ has been used. It is also worth
pointing out that both the eccentricity defined in Eq.~\rf{ecc} and the
oblateness defined in Eq.~\rf{obl} characterize the deviation of a spheroid
from a sphere. For a uniform spheroid, they are simply related via
\bea
\ep =  \frac{e^2}{2-e^2}.
\eea

%---------------------------------------------------------------------

%---------------------------------------------------------------------
\bibliography{refs}
%---------------------------------------------------------------------

\end{document}